\documentclass[11pt]{article} 
\usepackage{amssymb}
\usepackage{graphicx}
\usepackage{theorem}
\usepackage{url}
\newcommand \be{\begin{equation}}
\newcommand \bea{\begin{eqnarray}}
\newcommand \ee{\end{equation}}
\newcommand \eea{\end{eqnarray}}

\newcommand{\E}{{\rm E}}
\newcommand{\Var}{{\rm Var}~}
\newcommand{\Cov}{{\rm Cov}~}
\renewcommand{\(}{\left(}
\renewcommand{\)}{\right)}
\renewcommand{\[}{\left[}
\renewcommand{\]}{\right]}
\renewcommand{\epsilon}{\varepsilon}
\theoremheaderfont{\scshape}
\theorembodyfont{\upshape}

\begin{document}

\title{Self-Consistent Asset Pricing Models
\thanks{The authors acknowledge helpful discussions and 
exchanges with R. Roll. All remaining errors are ours.}}
\thispagestyle{empty}

\author{Y. Malevergne$^{1,2}$ and D. Sornette$^{3,4}$\\
\\
$^1$ ISFA Graduate School of Actuarial Science\\
University of Lyon 1, 50 avenue Tony Garnier,
69366 Lyon Cedex 07, France\\
$^2$ EM-Lyon Graduate School of Management\\
23 avenue Guy de Collongue, 69134 Ecully Cedex, France\\
$^3$ Department of Management, Technology and Economics\\
ETH Zurich, Kreuzplatz 5, CH-8032 Zurich, Switzerland\\
$^4$ LPMC, CNRS UMR6622 and Universit\'e des Sciences\\
Parc Valrose, 06108 Nice Cedex 2, France
\\
e-mails: Malevergne@univ-lyon1.fr and
dsornette@ethz.ch
}
\date{\today}
\maketitle

\abstract{We discuss the
foundations of factor or regression models in the light of the
self-consistency condition that the market portfolio (and more generally
the risk factors) is (are) constituted of
the assets whose returns it is (they are) supposed to explain.
As already reported in several articles, self-consistency
implies correlations between the return disturbances. As a consequence, 
the alpha's and beta's
of the factor model are unobservable. Self-consistency leads to 
renormalized beta's
with zero effective alpha's, which are observable with standard OLS regressions.
When the conditions derived from internal consistency
are not met, the model is necessarily incomplete, which means that some
sources of risk cannot be replicated (or hedged) by a portfolio of
stocks traded on the market, even for infinite economies.
Analytical derivations and numerical simulations show that, for arbitrary choices of
the proxy which are different from the true
market portfolio, a modified linear regression holds with a non-zero value
$\alpha_i$ at the origin
between an asset $i$'s return and the proxy's return. Self-consistency also
introduces ``orthogonality'' and ``normality'' conditions
linking the beta's, alpha's (as well as the residuals) and the weights of the proxy portfolio.
Two diagnostics based on these orthogonality
and normality conditions are implemented on a basket of 323 assets which have been
components of the S\&P500 in the period from Jan. 1990 to Feb. 2005.
These two diagnostics show interesting departures from dynamical self-consistency
starting about 2 years before the end of the Internet bubble.
Assuming that the CAPM holds with the self-consistency
condition, the OLS method automatically obeys the 
resulting orthogonality and normality conditions and therefore
provides a simple way to self-consistently assess
the parameters of the model by using proxy portfolios made only of the
assets which are used in the CAPM regressions. Finally,
the factor decomposition with the self-consistency condition
derives a risk-factor decomposition in the multi-factor case which is identical to the
principal components analysis (PCA), thus providing 
a direct link between model-driven and data-driven
constructions of risk factors.
This correspondence shows that PCA will therefore suffer
from the same limitations as the CAPM and its multi-factor generalization,
namely lack of out-of-sample explanatory power and predictability.
In the multi-period context, the self-consistency conditions force
the beta's to be time-dependent with specific constraints.}

\section{Introduction}

One of the most important achievements in financial economics is
the Capital Asset Pricing Model (CAPM), which is probably
still the most widely used approach to relative asset valuation. Its key
idea is that the expected excess return of an asset is proportional to
the expected covariance of the excess return of this asset with the
excess return of the market portfolio. The proportionality coefficient
measures the average relative risk aversion of investors.
As a consequence, there is an irreducible risk
component which cannot be diversified away, which cannot be
eliminated through portfolio aggregation and thus has to be priced.
The central testable implication of the CAPM is that assets
must be priced so that the market portfolio is mean-variance
efficient (Fama and French, 2004; Roll, 1977). However, past and recent tests 
have rejected the CAPM
as a valid model of financial valuation. In particular, the
Fama/French analysis
(Fama and French, 1992; 1993)
shows basically no support for the CAPM's central
result of a positive relation between expected return and global market
risk (quantified by the beta parameter). In contrast, other
variables, such as the market capitalization and the book-to-market ratio or
the turnover and the past return, present some explanatory power.

More and more sophisticated extensions of the CAPM beyond the
mean-variance approach
have not improved the ability of the CAPM and its generalization
to explain relative asset valuations. Let us mention the
multi-moments CAPM, which has originally been proposed by Rubinstein (1973)
and Krauss and Litzenberger (1976) to account for the
departure of the returns distributions from Normality. The relevance of
this class of models has been underlined by  Lim (1989) and Harvey
and Siddique (2000) who have tested the role of the asymmetry in
the risk premium by accounting for the skewness of the distribution of
returns and more recently by Fang and Lai (1997) and Hwang and
Satchell (1999) who have introduced a four-moments CAPM to take
into account the letpokurtic behavior of the assets return
distributions. Many other extensions have been presented such as the VaR-CAPM
(Alexander and Baptista, 2002), the Distributional-CAPM
(Polimenis, 2005), and generalized CAPM models
with consistent measures of risks and heterogeneous agents
(Malevergne and Sornette, 2006a), in order to account
more carefully for the risk perception of investors.

The arbitrage pricing theory (APT) provides an alternative to the CAPM.
Like the CAPM, the APT assumes that only non-diversifiable risk is
priced. But, unlike the CAPM which specifies returns as a
linear function of only systematic risk, the APT is based on the well-known
observations that multiple factors affect the observed time series of
returns, such as industry factors, interest rates, exchange rates, real
output, the money supply, aggregate consumption, investors confidence,
oil prices, and many other variables (Ross, 1976; Roll and Ross, 
1984; Roll, 1994).
While observed asset
prices respond to a wide variety of factors, there is much weaker
evidence that equities with larger sensitivity to some factors give
higher returns, as the APT requires.
This weakness in the APT has led to further generalizations of factor
models, such as the empirical Fama/French three factor model (Fama and French, 1995),
which does not use an arbitrage condition anymore. Fama and French
started with the observation that two classes of stocks show better
returns that the average market: (1) stocks with small market
capitalization (``small caps'') and (2) stocks with a high
book-value-to-price ratio (often ``value'' stocks as opposed to
``growth'' stocks).

What then survive of the fundamental ideas underlying the CAPM? A key
remark is that, given a set of assets, what is literally tested is the
efficiency of a specific proxy for the market portfolio together with
the CAPM.  As recalled by Fama and French (2004), the CAPM requires
using the market portfolio of all the invested wealth (which includes
stocks, bonds, real-estate, commodities, etc.). More precisely, as first
stressed by Roll (1977), ``The theory is not testable unless the exact
composition of the true market portfolio is known and used in the tests.
This implies that the theory is not testable unless {\it all} individual
assets are included in the sample.'' (italics in Roll (1977)).
Unfortunately, the market proxies used in empirical work are almost
always restricted to common stocks, and as pointed out by Roll, the
composition of a proxy for the market portfolio can cause quite
confusing inferences on the validity of the test and the mean-variance
efficiency of the market portfolio. It is thus
possible that the CAPM holds, the true market portfolio is efficient,
and empirical contradictions of the CAPM are due to bad proxies for the
market portfolio. Given a universe of $N$ assets, it is always possible
to construct a mean-variance portfolio (or any multi-moment
generalization thereof), which will be such that the expected excess
return of an asset is proportional to the expected covariance of the
excess return of this asset with the excess return of the mean-variance
portfolio. This results mechanically (or algebraically) from the
construction of the mean-variance portfolio. While this property
looks identical to the central test of the CAPM, in order for
the CAPM to hold and
for such a mean-variance portfolio to be the market portfolio, it should
remain a mean-variance portfolio ex-ante (out-of-sample). The failure
of the CAPM together with such a construction for the proxy of the
market portfolio is revealed by the notorious instability of mean-variance
portfolios (see for instance Michaud, 2003) with their weights needing to be continuously
readjusted as a function of time. Empirically, the problem is that a
mean-variance portfolio constructed over a given time interval will be no
more in general a mean-variance portfolio (even allowing for a
different average return) in the next period, and can not thus
qualify as the market portfolio.

In addition to this problem of the market portfolio proxy, the
``disturbances'' in factor models are correlated, as a consequence of
the self-consistency condition that, in a complete market, the market
portfolio and, more generally, the explanatory factors are made of (or
can be replicated by) the assets they are intended to explain
(Fama, 1973) (see also Sharpe (1990)'s Nobel
lecture). This presence of correlations between return residuals may
{\it a priori} pose problems in the pricing of portfolio risks: only
when the return residuals can be averaged out by diversification can one
conclude that the only non-diversifiable risk of a portfolio is born by
the contribution of the market portfolio which is weighted by the beta
of the portfolio under consideration. Previous authors have suggested
that this is indeed what happens in economies in the limit of a large
market $N \to \infty$, for which the correlations between residuals
vanish asymptotically and the self-consistency condition seems
irrelevant. For example, while Sharpe (1990; footnote 13)
concluded that, as a consequence of the self-consistency condition, at
least two of the residuals, say $\epsilon_i$ and $\epsilon_j$, must be
negatively correlated, he suggested that this problem may disappear in
economies with infinitely many securities. In fact, we show in
Malevergne and Sornette (2006b) that this apparently quite reasonable
line of reasoning does not tell the whole story: even for economies with
infinitely many securities, when the companies exhibit a large
distribution of sizes as they do in reality, the self-consistency
condition leads to the important consequence that the risk born out by
an investor holding a well-diversified portfolio does not reduce to the
market risk in the limit of a very large portfolio, as usually believed.
A significant proportion of ``specific risk'' may remain which cannot be
diversified away by a simple aggregation of a very large number of
assets. Moreover, this non-diversifiable risk can be accounted for in the APT
by an additional factor associated with the self-consistency condition.

Here, our more modest goal is to present a review of the foundation of factor models
using the self-consistent condition as a pivot to organize the presentation
and form threads across different results scattered in the literature.
Our goal will be reached if the reader starts to appreciate, as the authors did
in the course of their digestion of the literature leading to some
new results reported in (Malevergne and Sornette, 2006b), the many subtle
issues interconnecting the concepts of equilibrium, no-arbitrage and risk pricing.
In the physicist language, 
these concepts describe ultimately what can probably be seen as the attractive fixed point
(equilibrium) of self-organizing systems with feedbacks. We believe that
the study of the inner-consistency of these models can be useful to 
inspire the development of novel approaches addressing the above issues
and others.

The organization of the paper is the following. In the next section, we
consider an equilibrium model where the assets return dynamics can be
explained by a single factor, the market. At
equilibrium, this model is consistent with the CAPM but, due to the
self-consistency condition that the market portfolio is constituted of
the assets whose returns it is supposed to explain, the parameters of
the original factor model remain unobservable. Only the CAPM beta's are
observable if the true market portfolio is known. Due
the self-consistency condition, the residuals of the regression of the
assets' returns with respect to the market portfolio can only be defined
with a zero intercept.
Then, the orthogonality condition obtained in Fama (1973) concerning
the disturbances of the factor models
is derived both for a one-factor as well as for a multi-factor model.
In section 3, we discuss the calibration issues associated with the one factor model in
relation with the impact of the non-observability of the actual market
factor. We illustrate that, if a proxy is used (which is the real-life
situation), then one can only measure a modified beta value which may
differ from the true beta. In addition, a non-zero `alpha' appears, which has however
nothing to do with the unobservable alpha of the original factor model,
but reflects the difference between the proxy and the market portfolio.
Section 4 addresses the same question for multi-factor models. 
A multi-factor analysis with the self-consistency condition is shown
to be equivalent to the principal component analysis (PCA) applied to
baskets of assets. In the light of
these results, section 5 offers a discussion of the theoretical and
practical limitations of the factor-models. It underlines the necessity
for the introduction of non constant $\beta$'s and propose some
restrictions on the possible dynamics for the $\beta$. All the 
technical derivations
are gathered in the 6 appendices.

\section{Self-consistency of factor models}

\subsection{One-factor model: dynamical consistency of the CAPM}
\label{sec:capm}

\subsubsection{Factor model from CAPM}

The celebrated Capital Asset Pricing Model, derived by Sharpe (1964),
yields the famous relation known as the {\em Market Security Line}
\be
\E\[r_i\] = r_0 + \beta_i \cdot \E\[r_m -r_0\],
\ee
where $r_m$, $r_i$ and $r_0$ denote the market return, the
return\footnote{Given the price $P_i(t)$ of security $i$ at time $t$, is
return is defined as $r_i(t)=\frac{P_i(t+1)}{P_i(t)}-1$.} on asset $i$
and the risk free interest rate respectively, while
\be
\beta_i = \frac{\Cov \(r_i, r_m \)}{\Var r_m}.
\ee
As stressed by Sharpe (1990), ``the value $\beta_i$ can be given an
interpretation similar to that found in regression analysis utilizing
historic data, although in the context of the CAPM it is to be
interpreted strictly as an {\it ex-ante} value based on probabilistic
beliefs about future outcomes.'' If the investors' anticipations are
self-fulfilling, the
relationship between $r_i$ and $r_m$ can be modeled as
\be
r_i = a_i + \beta_i \cdot r_m + \epsilon_i~,
\label{mnsjlkl.s}
\ee
with $a_i=\(1-\beta_i\)r_0$, provided that the expectation of the
residual $\E\[\epsilon_i\]$ is assumed to be zero. These two 
conditions $a_i=\(1-\beta_i\)r_0$ and $\E\[\epsilon_i\]=0$
ensures that the market portfolio is efficient in the mean-variance sense.
Indeed, taking expectations (or sample
means) of (\ref{mnsjlkl.s}), one obtains an exact linear cross-sectional relation between
mean returns and beta's.  There is a one-to-one correspondence between
exact linearity and mean/variance efficiency of the market portfolio (Bodie et al., 2004).

\subsubsection{CAPM from a factor model}

Let us now start from the opposite view point to determine
the conditions under which the CAPM relation holds for an economy 
obeying a linear factor model, where the excess returns of asset prices
over the risk-free rate $r_0$ are determined according to the following
equation \footnote{in all what follows, we work with excess returns, i.e.,
returns decreased by the risk-free rate $r_0$ but use the same notation
as for the returns to simplify the notations.}
\be
\label{eq:capm}
\vec r_t = \vec \alpha + \vec \beta^0 \cdot r_m(t) + \vec \epsilon_t,
\ee
where $\vec r_t$ is the $N \times 1$ vector of asset excess returns at time $t$,
$r_m(t)$ is the excess return on the market portfolio and $\vec \epsilon_t$ is a
vector of disturbances with zero average $\E \[\vec \epsilon_t\] = \vec 0$ and
covariance matrix $\Omega_t = \E \[ \vec \epsilon_t \cdot \vec \epsilon_t'\]$.
We assume that $\Omega_t$ is a deterministic function of $t$ and that the
$\vec \epsilon_t$ are independent. We do not make any other assumption
concerning $\Omega_t$, in particular, we do not assume that it is a
diagonal matrix since the CAPM places no restriction on the correlation 
between the disturbance terms. The symbols $\vec \alpha$ and $\vec \beta^0$ 
represent constant $N \times 1$ vectors.

Let us assume that the model (\ref{eq:capm}) is common knowledge, {\it
i.e.}, each economic agent knows that the asset returns follow equation
(\ref{eq:capm}), each agent knows that all other agents know that the
assets returns follow equation (\ref{eq:capm}), and so on... Let
us assume that, by reallocating her wealth $W_t$ among the $n$ risky assets and
the risk-free asset at each intermediate time period $t=1, \ldots, T-1$,
each agent aims at maximizing her expected terminal wealth $W_T$
under the constraint that its variance $\Var W_T$
is not greater than a predetermined level $\sigma_{W_T}^2$.
Mathematically, this dynamic optimization program reads
\bea
&&\quad~~ \max_{\vec w} \E \[W_T\]\nonumber\\
({\cal P}):\qquad &&{\rm s.t}~ \Var W_T \le \sigma_{W_T}^2\\
&& \quad~~ W_{t+1} = W_t \[1+ \vec w' \vec r_t + r_0\], \quad t=0, 1, 
\ldots, T-1.\nonumber
\eea
Many other approaches have been considered in the large body of
literature devoted to the problem of optimal investment selection in a
multi-period framework. In particular, the approaches based on the
maximization of the expected utility of the terminal wealth or of the
lifetime consumption seem to dominate, but they often rely on a
specific choice of the utility function, such as the CARA, HARA or
quadratic utility functions (Samuelson, 1969, Hakansson 1971, Pliska
1997, among many others). Since the choice of a particular utility
function may appear as arbitrary, we have preferred to resort to the
mean-variance criterion in so far as it constitutes a low order expansion
approximation which holds irrespective of the specific form of the
utility function.

The solution of problem $({\cal P})$ can be found for instance in Li and Ng (2000):
at each time period $t$, the optimal strategy amounts to
invest a fraction of wealth in the risk free asset and the remaining in
the risky portfolio
\be
\label{eq:dghpo}
\vec w_t^* = \frac{{\Sigma_t}^{-1} \E\[ \vec r_t \]}{\vec 1' 
{\Sigma_t}^{-1} \E\[ \vec r_t \]},
\ee
where $\Sigma_t = \Cov \vec r_t$ denotes the covariance of the vector of excess returns
of the asset prices over the risk-free rate, at time $t$. As we shall
see in the sequel, $\Sigma_t$ and $\E\[\vec r_t\]$ are known
functions of $t$, which is a necessary assumption for the solution given
by Li and Ng (2000) to hold. 

Since all agents invest only in two funds, namely the risk-free asset
and the risky portfolio with weights $\vec w^*_t$, if we assume that 
an equilibrium
is reached at each time $t$, then the composition $\vec w^*_t$ of the risky
portfolio must represent that of the market portfolio at time $t$. 
In other words, in full generality, $\vec w_t^*$ given by
(\ref{eq:dghpo}) is nothing but the efficient tangency portfolio on the
frontier composed of the existing risky assets.  It becomes the market
portfolio of all assets when the assets being considered here comprise
indeed all assets, which is the case we first examine. Section
\ref{sec:fgm} discusses what happens when this is not the case.
For the sake of simplicity, we will denote
by $\vec w_t$ the composition of the market portfolio.

It is important to note that the result (\ref{eq:dghpo}) 
holds irrespective of the time horizon
$T$ chosen by the investors because the composition $\vec w_t$ of the market
portfolio is independent of $T$. Only the relative part of
wealth invested in the risk-free asset and in the market portfolio
depends on $T$, but this has no effect on the composition $\vec w_t$ of the market
portfolio. As a consequence,
the result still holds when investors have different time horizons,
as in real markets. 

Now, accounting for the fact that the market factor is itself built upon
the universe of assets that it is supposed to explain (which we
refer to as the ``self-consistent condition''), the model must
fulfill the internal consistency condition
\be
r_m(t) = \vec w_t' \cdot \vec r_t.
\label{mvgmlsl}
\ee
Starting from this self-consistency condition together with the
assumption that investors follow a dynamic mean-variance strategy and with
the condition of market equilibrium, we show in Appendix A that the
regression model (\ref{eq:capm}) leads to the CAPM
\be
\label{eq:pjzer2}
\E \[ \vec r_t\] = \vec \beta_t \E\[r_m(t)\],
\ee
with
\be
\label{eq:threj2}
\vec \beta_t  = \frac{\Cov \(\vec r_t, r_m(t) \)}{\Var~ r_m(t)} =
\frac{\(\vec 1 - \vec \beta^0\)' \Omega_t^{-1} \vec \alpha}{\vec
\alpha' \Omega_t^{-1} \vec \alpha}~ \vec \alpha + \vec \beta^0~.
\ee
This shows that the regression model (\ref{eq:capm}) is consistent with
the relation of the CAPM provided that the internal consistency
condition (\ref{mvgmlsl}) holds together with the existence of an
equilibrium.

The rather lengthly derivation in Appendix A is not needed in
the standard approach in which the vector $\vec \alpha$ is identically
zero and the market portfolio is mean-variance efficient as given
by (\ref{eq:dghpo}). Appendix A makes explicit that the parameters of the
market model (\ref{eq:capm}) are of no consequence for the CAPM.  Appendix A derives the
expression of the observable parameters of the CAPM (in particular the
beta) from the parameters $\alpha$'s, $\beta_0$'s and the matrix 
$\Omega$ of the covariance
of the disturbances $\vec \epsilon$ of the market model\footnote{As
we clarify further below, the disturbances $\vec \epsilon$ 
of the market model are not the residuals of an OLS
(ordinary least-square) regression}. 

Therefore, the general regression model (\ref{eq:capm})
provides a reasonable statistical model to test the CAPM
relation (\ref{eq:pjzer2}). But, two important point must be discussed.
First, even if $\vec \alpha$ and $\vec \beta^0$ are assumed constant,
the CAPM's $\beta$ depends on time $t$ as soon as $\Omega_t$ is not
constant. Thus, the heteroscedasticity of the residuals is sufficient to
make the $\beta$'s time varying. Since, in the real market, the variance
of assets returns is time varying (the so-called GARCH effect), one has
to account for the dynamics of the $\beta$'s. Second, the equilibrium
imposes a dynamic constraint on the composition of the market portfolio.
On the one hand, it is endogenously determined by the investors'
anticipations according to formula (\ref{eq:dghpo}). On the other
hand, the market portfolio must be related to the market capitalization
of each asset, which reflects the economic performance of the firms.
Thus, the relation
\be
\label{eq:dfjh}
w_{t+1}^i = w_t^i \cdot \frac{1+r_t^i + r_0}{1+r_m(t) + r_0}
\ee
must hold. The $r_0$ appears in the numerator and denominator because of
our convention to denote by $r_t^i$ and $r_m(t)$ the excess returns of
asset and market prices over the risk-free interest $r_0$. For the time
being, we assume that this relation (\ref{eq:dfjh}) is compatible with
the dynamics described by (\ref{eq:capm}) and with the optimal portfolio allocation
(\ref{eq:dghpo}) and will discuss this point in more detail at the end
of this article.

\subsection{One-factor model: observable parameters, orthogonality
and normalization conditions \label{mgvmlaa}}

For ease of the exposition, let us
assume that $\Omega_t$ remains constant during the time interval under
consideration. As a consequence, $\vec \beta$ can be {\it a priori}
independent of $t$ as shown by eq. (\ref{eq:threj2}), allowing us to
remove the subscript $t$ in the sequel. 

The previous sub-section has made clear that, according to
(\ref{eq:threj2}), the coefficients $\vec \beta$ of the CAPM can be
expressed in terms of the $\alpha$'s, $\beta_0$'s and the matrix
$\Omega$ of the covariance of the disturbances $\vec \epsilon$ of the
market model. Actually, one can go further and show that 
the self-consistency condition implies that only $\vec \beta_t$ is
observable while the coefficients $\vec \alpha$ and $\vec \beta^0$ are
unobservable. Indeed, expression (\ref{eq:capm}) cannot be directly
calibrated by the OLS estimator since the disturbances $\vec \epsilon_t$
are correlated with the regressors while an OLS estimation automatically
construct residuals which are orthogonal to the factor decomposition.
To see why the disturbances $\vec \epsilon_t$
are correlated with the regressors $r_m(t)$, let us left-multiply
expression (\ref{eq:capm}) by $\vec w_t'$. Then, the self-consistency condition
(\ref{mvgmlsl}) implies that
\be
\label{eq:market_bis}
r_m(t)= \frac{{\vec w'}_t \cdot \(\vec \alpha + \vec \epsilon_t \)}{1 -
{\vec w'}_t \cdot {\vec \beta}^0}~,
\ee
unless $w_t \vec \beta^0 = 1$. 

The fact that the regressors $r_m(t)$ are correlated with the residuals $\vec \epsilon_t$
does not invalidate the OLS procedure. It just means that the OLS procedure
will estimate residuals which are different from the model disturbances.
The observed residuals are obtained by decomposing 
the disturbances $\vec \epsilon_t$ on its component correlated with
$r_m(t)$ plus a contribution uncorrelated with $r_m(t)$.
We thus introduce two non-random vectors $\vec \delta $, $\vec \gamma$ and
the random vector $\vec u_t$, {\em uncorrelated} with $r_m(t)$ with
zero mean, such that
\be
\label{eq:rtyho}
\vec \epsilon_t = \vec \delta + \vec \gamma \cdot r_m(t)  + \vec u_t.
\ee
Then, Appendix B shows that the one-factor model reduces to
\be
\vec r_t = \vec \beta \cdot r_m(t) + \vec u_t~,
\label{eq:eriugh2}
\ee
with the ``normalization'' and ''orthogonality'' conditions
\be
\vec w_t' \vec \beta =1~ \quad {\rm and} \quad  \vec w_t' \vec u_t = 0~,
\label{mvmlsl}
\ee
which derive from the self-consistency condition (\ref{mvgmlsl}). The result
(\ref{eq:eriugh2}) means that, {\em under the assumption that $r_m(t)$ is
observable}, the OLS estimator of (\ref{eq:capm}) provides an estimate
of $\vec \beta$ {\em and not of $\vec \beta^0$ and $\vec \alpha$ which
remain unobservable}. Taking the expectation of (\ref{eq:eriugh2}) recovers
the CAPM prediction (\ref{eq:pjzer2}) as it should.

We should stress that the orthogonality condition $\vec w_t' \vec u_t =
0$ shows that at least two of the $u_{t,i}$ must be negatively
correlated, which resemble Sharpe (1990)'s statement in his footnote 13.
But, there is an important difference in that the
regression (\ref{eq:eriugh2}) has zero intercept (its ``alpha'' is zero).
The absence of intercept
together with the mean-variance nature of the market portfolio
automatically ensures the validity of the CAPM relation (\ref{eq:pjzer2}).

Using the jargon of physicists, we can rephrase these results as
follows. The self-consistency condition together with the mean-variance
efficient nature of the market portfolio imply that the market model
(\ref{eq:capm}) is ``renormalized'' into an observable model given by
expression (\ref{eq:eriugh2}) with (\ref{mvmlsl}), that is, the ``bare''
parameters  $\vec \alpha$ and ${\vec \beta}_0$ are renormalized into
$\vec 0$ and $\vec \beta$. A standard OLS regression (a measurement)
gives access only
to the renormalized values $\vec 0$ and $\vec \beta$, in the same that
physicists can only measure for instance the large scale renormalized mass and
charge of an electron and not its bare values (Lifshitz et al., 1982).

\subsection{Multi-factor model \label{mgmsa;a}}

Let us generalize (\ref{eq:capm}) and
assume that the excess return vector $\vec r_t$ of $n$ securities traded
on the market (made of these $n$ assets), over the risk free interest rate,
can be explained by the $q$-factor model
\bea
\vec r_t &=& \sum_{i=1}^q \vec \beta_i u_i(t) + \vec \epsilon_t, \\
&=& B \vec u(t) + \vec \epsilon_t,
\label{eq:kukggf}
\eea
where $B$ is the $n \times q$ matrix which stacks the vectors $\vec
\beta_i$, $\vec u_t$ is the vector whose i$^{th}$ component is the
i$^{th}$ risk factor $u_i$ and $\E \[\vec \epsilon(t) \]=0$.

With $n$ assets and $n+q$ sources of randomness, the
market is {\it a priori} incomplete. The market becomes complete
if all risk factors can be replicated by an asset portfolio.

Consider the risk factor $i$, which can be replicated by the portfolio
$\vec w_i$, that is, $u_i(t) = \vec w_i' \vec r_t$ in vector notations. The
internal consistency of the model implies that
\be
\vec w_i' \vec r_t = u_i(t)
=\sum_{j=1}^q \( \vec w_i' \vec \beta_j\) u_j(t) + \vec w_i' \vec \epsilon_t,
\label{eq:PRTI}
\ee
so that
\be
\sum_{j \neq i} \( \vec w_i' \vec \beta_j\) u_j(t) + \( \vec w_i' \vec
\beta_i - 1\) u_i(t) + \vec w_i' \vec \epsilon_t=0~.
\label{mmzla;q}
\ee

For a complete market such that all the risk factors $u_i$'s can be
replicated by asset
portfolios $\vec w_i$'s, $i=1,\ldots,q$ and denoting by $W$ the matrix
which stacks all the portfolio weight vectors $\vec w_i$'s, the
self-consistency condition (\ref{mmzla;q}) generalizes into
\be
\label{eq:jhghf}
\( {\rm Id} - W'B\) \vec u(t) = W' \vec \epsilon_t~.
\ee
Taking the expectation of both sides yields
\be
\( {\rm Id} - W'B\) \E\[\vec u(t)\] = 0~,
\ee
since we assume $\E \[\vec \epsilon(t) \]=0$. Two cases
must be considered.
\begin{itemize}
\item First case: $\det \( {\rm Id} - W'B\) \neq 0$ and the unique solution
is $\E\[\vec u(t)\] = 0$, so that $\E \[ \vec r_t \] = 0$ by
(\ref{eq:kukggf}), which does not capture a real economy.

\item Second case: $\det \( {\rm Id} - W'B\) = 0$, which means that
the matrix $W'B$
has rank $q-p$, for some $0<p \leq q$. Provided
that the system admits a solution, this solution can be expressed as a linear
combination of $p$ independent vectors. As a consequence, the expected
excess return on each individual asset $\E\[r_i\]$ can be expressed as the
linear combination of the expected value of only $p$ risk factors.
Therefore, only $p$ factors really matter. This implies that, if we assume
that assets excess returns really depend upon $p=q$ factors, the rank of the
matrix $\( {\rm Id} - W'B\)$ should be $q-p=0$ so that the expectation of
the excess return on each individual asset $\E\[ r_i\]$ can be expressed as the
linear combination of the expected value of all the $q$ risk factors. In
such a case, we will say that the model is irreducible, an hypothesis that
we will assume to hold in the sequel. The case $p<q$ can be treated
analogously by expressing the excess return of each individual asset as a
linear combination of the expected value of the $p$ risk factors.
\end{itemize}

The condition that the rank of the matrix $\( {\rm Id} - W'B\)$ should be zero
for the asset excess returns to depend on the $q$ irreducible factors simply
means that the normalization condition
\be
\label{eq:sdfkuh}
W'B = {\rm Id}
\ee
must hold. This relation is satisfied by the market factor in the CAPM,
and generalizes the normalization condition discussed in section 
\ref{sec:capm}.
In addition, equation (\ref{eq:jhghf}) together with (\ref{eq:sdfkuh}) enables
us to conclude that
\be
\label{eq:pgho}
W' \vec \epsilon_t = 0~,
\ee
which means that the vector $\vec \epsilon_t$ of disturbances has dimension
$n-q$ at most, provided that $W$ is full rank, {\em i.e.} provided that
the $q$ risk factors $u_i(t)$ can be replicated by $q$ linearly
independent portfolios $\vec w_i$. Condition (\ref{eq:pgho})
generalizes the orthogonality condition for the one-factor model
derive in section \ref{mgvmlaa}. The two conditions
(\ref{eq:sdfkuh}) and (\ref{eq:pgho}) generalize the orthogonality 
and normalization
conditions (\ref{mvmlsl}) obtained for the one-factor CAPM.

Note that $\vec u$ and $\vec \epsilon$ are uncorrelated under the condition
that the $q$ risk factors $u_i(t)$ can be
replicated by $q$ linearly
independent portfolios.

To sum up, the possibility to replicate the risk factors by portfolios implies
strong internal consistency conditions for factor models, namely
equations (\ref{eq:sdfkuh}) and (\ref{eq:pgho}). Conversely, if these
conditions are not met, the model is necessarily incomplete, which means
that some sources of risk cannot be replicated (or hedged) by an asset
portfolio. Therefore, risk factors, such as the GDP, the term spread,
the dividend yield, the size and book-to-market factors (Fama and
French, 1993; 1995)
and so on, could bring in additional information with respect to
the usual market factor. See Petkova (2006) for empirical evidence.

\section{Non-observability of the market portfolio (One-factor model)}
\label{sec:fgm}

\subsection{What if the proxy is different from the true market portfolio?
\label{mmslsl;}}

In practice, the true market factor is unknown and one commonly uses a
proxy. We show in Appendix C that model (\ref{eq:eriugh2}) leads to
\be
\vec r_t = \underbrace{\( \E\[r_m\] \vec \beta -  \E\[\tilde r_t\] \vec
{\tilde \beta}\)}_{\vec{\tilde \alpha}}
+ \vec {\tilde \beta} \cdot \tilde r_t + \vec \eta_t~,
\label{mgmlss}
\ee
where $\tilde r_t$ is the proxy excess return, $\vec {\tilde \beta}$ is the vector
of beta's of the regression of asset excess returns on the proxy and $\vec \eta_t$
has zero mean $\E\[\vec \eta_t \]=0$ and is uncorrelated with the proxy
$\Cov\(\vec \eta ,\tilde r_t\) = 0$. The explicit
dependence of $\vec {\tilde \beta}$ as a function of the true $\vec \beta$,
the weights $\tilde w_t'$ of the portfolio proxy, the variance ${\rm Var} r_m$
of the market portfolio excess returns and the covariance matrix $\tilde \Omega$ of 
the vector $\vec u_t$ of residuals of the model (\ref{eq:eriugh2})
is given in equation (\ref{eq:dhml}).

The result (\ref{mgmlss})
derives straightforwardly from the CAPM formulated explicitly with
(\ref{eq:eriugh2}) and (\ref{mvmlsl}) by again using a self-consistent
(or endogenous) condition that the proxy is itself a portfolio of the assets
it is supposed to explain. As a consequence of the internal consistency
requirement, one gets new orthogonality and normalization conditions. As
previously, we have the normalization and orthogonality conditions
\be
\label{eq:msfjilj}
\tilde w_t' \vec{\tilde \beta} = 1, \quad {\rm and} \quad \tilde w_t' 
\vec \eta_t=0,
\ee
where $\tilde w_t$ represents the composition of the proxy at time $t$.
In addition, we have the following orthogonality constraint
\bea
\tilde w_t' \vec{\tilde \alpha} &=& \tilde w_t'\( \E\[r_m\] \vec 
\beta -  \E\[\tilde r_t\]
\vec {\tilde \beta}\), \nonumber\\
&=& \underbrace{\(\tilde w' \vec \beta\)}_{\beta~{\rm of~ the~ proxy}}
\cdot \E\[r_m\]  -  \E\[\tilde r_t\],\nonumber\\
&=& 0~,
\label{mmkvkmlala}
\eea
provided that the CAPM relation holds.

Using a proxy instead of the true market portfolio
yields a non-vanishing intercept $\vec{\tilde \alpha} =
\E\[r_m\] \vec \beta -  \E\[\tilde r_t\] \vec {\tilde \beta}$ 
in the regression of the excess returns of each asset
as a function of the excess returns of the portfolio proxy, which
is {\it a priori} different from asset to asset. 
However, taking the expectation of (\ref{mgmlss}), we obtain
\be
\E\[r_{i,t}\] = \E\[r_m\] \beta_i = \left({\E\[r_m\] \over \E\[\tilde r_t\]}
{\beta_i \over {\tilde \beta}_i}\right) \E\[\tilde r_t\]~\tilde \beta_i~,
\label{mbnsjkswls}
\ee
for each individual asset $i$. 
As in the standard CAPM prediction, we thus obtain that
the expected excess return $\E\[\vec r_{i,t}\]$
of an asset $i$ is proportional to its beta ${\tilde \beta}_i$ (obtained from the 
conditional regression (\ref{mgmlss})). But there is a major difference
with the standard CAPM prediction, which is that the coefficient of
proportionality is not simply the expectation $\E\[\tilde r_t\]$
of the proxy excess returns (as one could expect naively from translating
the standard result to the proxy case). The difference involves
the two correction factors $\E\[r_m\] / \E\[\tilde r_t\]$ and
$\beta_i / \tilde \beta_i$, the second one being non-constant since
it is a function of $\tilde \beta_i$ itself. Recall that $\E\[r_m\]$
and the $\beta_i$'s are in principle unobservable. We can thus 
expect a deviation from the standard CAPM linear relationship
due to an increased scatter induced by the scatter in the coefficient
of proportionality between expected excess return and beta evaluated with 
a market proxy.

Although this result is generally true, there is an exception.  If the proxy
happens to be on the ex-ante mean/variance efficient frontier, there
will be an exact cross-sectional relation between expected returns and
betas (calculated against the proxy) and there will be no scatter
around the linear relation between mean returns and beta's.  Any
market proxy will produce exact linearity, not just the tangency
portfolio from the translated (by $r_0$) origin.  Of course, the beta's will
be different for each such proxy but there will be no scatter. 
Generally, there is no need to assume the existence of a riskless rate. 
This is the heart of Black (1972)'s generalization of the CAPM.  If there is no
riskless rate, any ex-ante mean-variance efficient portfolio, which can
lie anywhere on the positive or negative part of the frontier, will
produce exact cross-sectional mean return/beta linearity.  The only
exception is the global minimum variance portfolio, which is positively
correlated with all assets.  For all other market proxies, there is a
``zero-beta'' portfolio, a portfolio uncorrelated with the chosen proxy,
which serves in place of the riskless rate.

\subsection{Empirical illustration}

\begin{figure}
\begin{center}
\includegraphics[width=14cm]{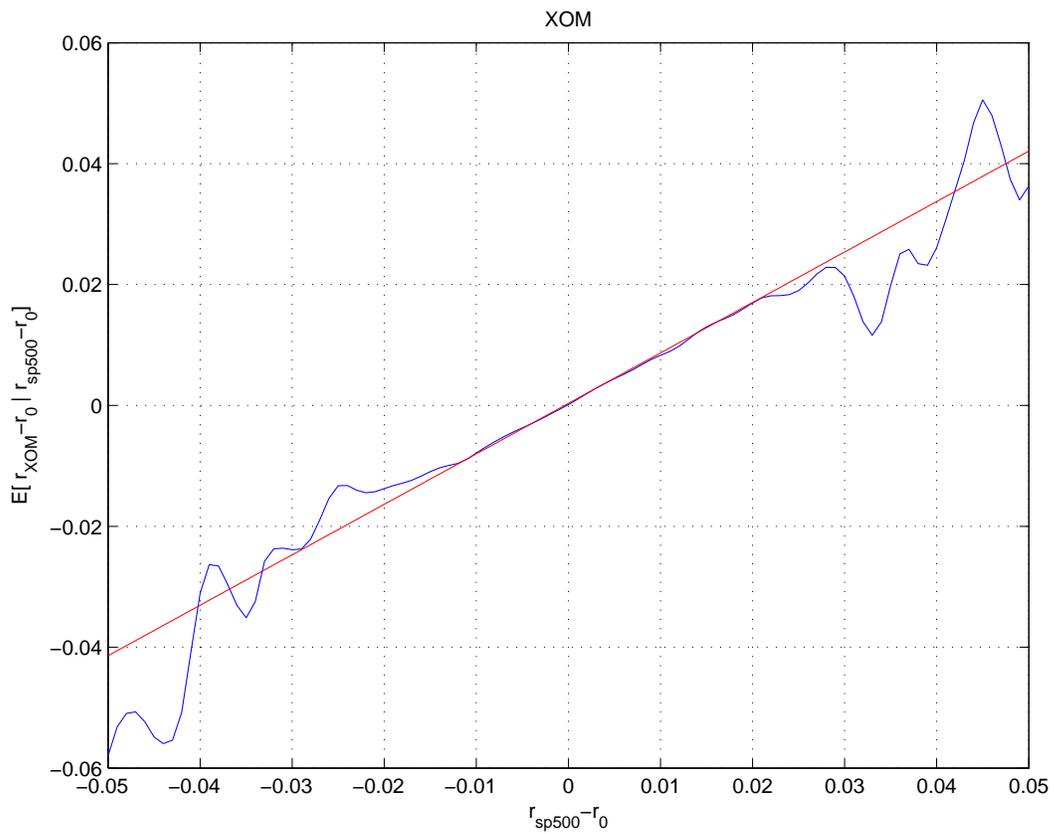}
\end{center}
\caption{Regression of the expected return above the
risk-free interest rate for Exxon mobil daily returns
with respect to the excess return of the S\&P500 index
over the period from July 1962 to December 2000. The risk free interest rate 
is obtained from the three
month Treasury Bill.}
\label{XOM}
\end{figure}

As an illustration, let us first take the S\&P500 index as a proxy for
the USA market portfolio. Figure \ref{XOM} shows the average daily return
of Exxon mobil (ticker XOM) daily returns conditioned on a fixed value of
the S\&P500 index daily returns $r_m(t)$
over the period from July 1962 to December 2000. In practice, 
we consider a given value $r_m$ (to within a small interval) of the S\&P500. We
then search for all days for which the return of the S\&P500 was equal to
this value $r_m$ (to within a small interval). We then take the average of
the daily return of Exxon mobil realized in all these days. We then
iterate by scanning all possible values of $r_m$ and use a kernel
estimation to get a smoother and more robust estimation. Note that
this procedure is non-parametric and
provides an interesting determination of the market model. Indeed,
suppose that the return $r_i$ of an asset $i$ is given by
\be
r_i(t) = F_i[r_m(t)] + e(t)~,
\ee
where $F_i[x]$ is an a priori arbitrary (possibly non-linear function) and
$e(t)$ are the zero-mean residuals.  Then, the above non-parametric procedure
(whose result is shown in figures 1 and 2) amounts to calculate ${\rm E}[r_i |
r_m=x]$ as a function of $x$:
\be
{\rm E}[r_i | r_m=x] = F_i[x]~.
\ee
Figure \ref{XOM} plots the function $F_i[x]$ determined non-parametrically
from the data. It seems that a linear dependence proves a reasonable
approximation of the data presented in Fig.~\ref{XOM}.
The straight line is the line
of equation $y = \alpha_{\rm XOM} + \beta_{\rm XOM} \cdot r_{SP500}(t)$,
where $\beta_{\rm XOM}$ is obtained
from the regression
\be
\label{eq:kjd}
r_{\rm XOM}(t) = \alpha_{\rm XOM} + \beta_{\rm XOM} \cdot
r_{SP500}(t) + \epsilon_{\rm XOM}(t)
\ee
of the returns. 

\begin{figure}
\begin{center}
\includegraphics[width=14cm]{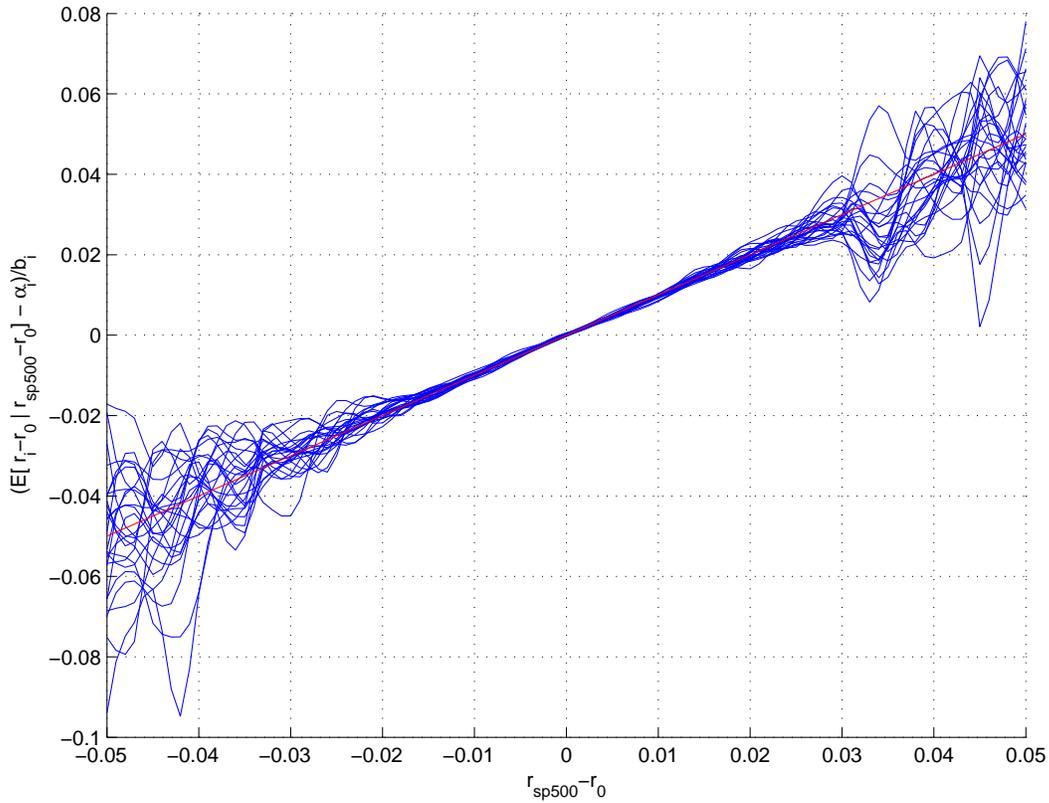}
\end{center}
\caption{Each curve is similar to that shown in figure \ref{XOM}
and represents the normalized expected return above the
risk-free interest rate defined by (\ref{mnvlaaa}) for a given stock $i$
over the period from July 1962 to December 2000 as a function of the 
excess return
$r_{SP500} - r_0$ above the risk-free interest rate $r_0$ of the S\&P500
index taken as a proxy of the market portfolio. Since the $\alpha_i$'s
and $\beta_i$'s are different from asset
to asset, the normalization (\ref{mnvlaaa}) ensures by construction that
a good linear regression for each asset should be qualified by having
all curves
collapse on the diagonal, with unit slope and crossing of the origin, as
observed up to statistical fluctuations. The 25 curves corresponds to
the following stocks: Abbott Laboratories, American Home Products Corp.,
Boeing Co., Bristol-Myers Squibb Co., Chevron Corp., Du Pont (E.I.) de
Nemours \& Co., Disney (Walt) Co., General Electric Co., General Motors
Corp., Hewlett-Packard Co., International Business Machines Co.,
Coca-Cola Co., Minnesota Mining \& MFG Co., Philip Morris Cos Inc.,
Merck \& Co Inc., Pepsico Inc., Pfizer Inc., Procter \& Gamble Co.,
Pharmacia Corp., Schering-Plough Corp., Texaco Inc., Texas Instruments
Inc., United Technologies Corp., Walgreen Co. and Exxon Mobil Co.
The risk free interest rate is obtained from the three
month Treasury Bill.}
\label{Renormalized}
\end{figure}

This plot presented in Fig.~\ref{XOM} is typical of the relationship between
conditional expected returns as a function of the return of the S\&P500 index,
obtained for all stocks in the S\&P500, as shown from the superposed data in
figure \ref{Renormalized}. Figure \ref{Renormalized} is the same as 
figure \ref{XOM}, but for 25 
different assets. In order to represent the corresponding functions
$F_i(x)$ for each asset on a same figure without loosing visibility, we
have just translated and scaled each curve, i.e., we plot 
\be
\frac{\E \[r_i - r_0 | r_{SP500} - r_0 \] - \alpha_i}{\beta_i} = (F_i[x]-\alpha_i)/\beta_i~,
\label{mnvlaaa}
\ee
as a function of $x=r_{SP500} - r_0$, where the $\alpha_i$'s and $\beta_i$'s 
are obtained by linear regressions similar to (\ref{eq:kjd}), one fit being performed
for each non-parametrically determined $F_i$. The risk-free
interest rate $r_0$ is basically negligible at the daily scale.
$\E \[r_i - r_0 | r_{SP500} - r_0 \]$ is the expected return of stock $i$
above the risk-free interest rate, conditional on the value of
$r_{SP500} - r_0$.
The straight line in Figure \ref{Renormalized}
has slope $1$ and goes
through the origin, thus confirming the remarkable quality of the
relationship between the conditional expected asset returns
and the S\&P500 index daily returns, in agreement with (\ref{mgmlss}).
In other words, Figure \ref{Renormalized} seems to confirm that the
$F_i$'s appear to be quite closely
approximated by an affine function: $F_i[x]=\alpha_i + \beta_i x$.

\begin{figure}
\includegraphics[width=14cm]{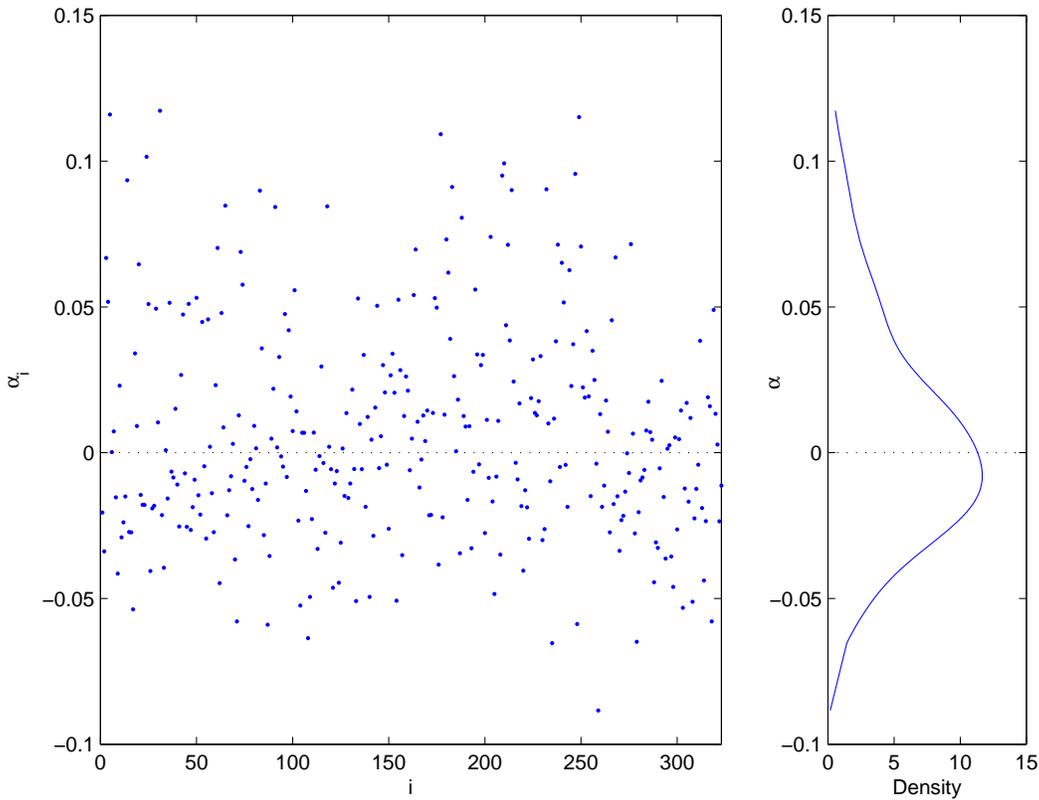}
\caption{Left panel: Population
of the intercepts of the regression of the expected monthly excess returns of
323 stocks entering into the composition of the S\&P500 between January
1990 and February 2005 versus the monthly excess returns of the 
effective S\&P323 index that we have constructed 
as a portfolio of these 323 stocks with weights proportional
to their capitalizations. The risk free interest rate is obtained from the three
month Treasury Bill. The abscissa is an
arbitrary indexing of the 323 assets. The estimated probability density
function of the population of alpha's is shown on the right panel
and illustrates the existence of a systematic bias for the alpha's.}
\label{InterceptRealData_ExcessReturnSP323}
\end{figure}

We have performed similar regressions as a function of the S\&P500
returns for the monthly returns of the 323 stocks which remained into the composition of
the S\&P500 over the period between January 1990 and February 2005.
But, in order to test the self-consistency condition and its consequences
derived above, one could argue that it should be better to construct a market portfolio based
solely on these 323 stocks. We have thus constructed an effective
S\&P323 index, constituted as a portfolio of these 323 stocks with
weights proportional to their capitalizations. The regressions of the
expected monthly returns of each of these 323 stocks conditioned on the
S\&P323 index monthly returns as a function of the S\&P323 index monthly
returns are similar to those obtained on the S\&P500 and resemble the
regressions shown in figures \ref{XOM} and \ref{Renormalized} albeit
with more noise (not shown). Figure
\ref{InterceptRealData_ExcessReturnSP323} shows the population of the
intercepts (the alpha's) of these regression. The abscissa is an
arbitrary indexing of the 323 assets. The estimated probability density
function of the population of alpha's is shown on the right panel and
illustrates the existence of a systematic bias for the alpha's, as
expected from the previous section \ref{mmslsl;}. Note that the bias is
negative which reflects the fact that over the period of study, the
average performance of the S\&P323 (and even more so for the
S\&P500) has been smaller than the risk-free rate. 
Another way of formulating the existence of the bias is just to say that
the constructed index is not located on the sample
efficient frontier.

\begin{figure}
\includegraphics[width=14cm]{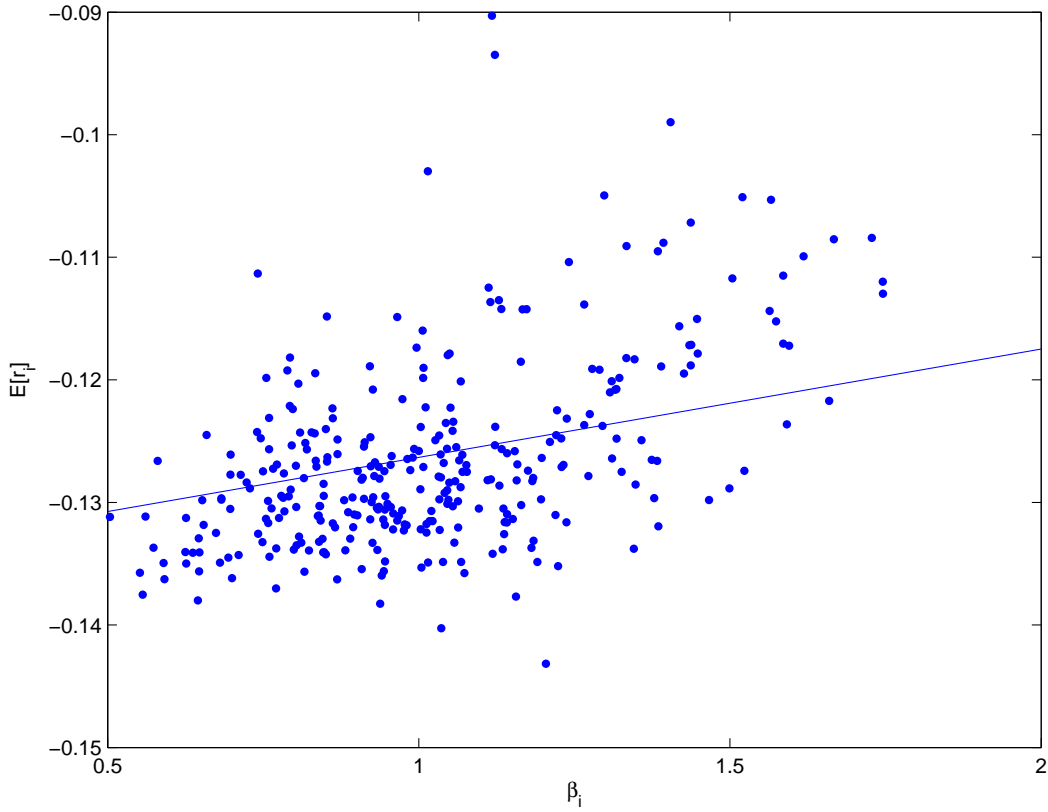}
\caption{Expectation $\E \[r_i -r_0\]$ of the monthly excess returns of the 323 assets
used in figure \ref{InterceptRealData_ExcessReturnSP323} as a function of their $\beta_i$
obtained by regressions with respect to the excess return to the effective
S\&P323 index. The risk free interest rate is obtained from the three
month Treasury Bill. Under the
CAPM hypothesis, one should obtain a straight line with slope $\E
\[r_{SP323} - r_0\]$ ($-13.1\%$ per month) and zero additive coefficient at the origin. The
straight line is the regression $y = 0.88\% -13.5\% \cdot x$.}
\label{CapmRealData_ExcessReturnSP323}
\end{figure}

Figure \ref{CapmRealData_ExcessReturnSP323} plots the expected returns $\E
\[r_i -r_0\]$ of the monthly excess returns of the 323 assets used in
figure \ref{InterceptRealData_ExcessReturnSP323} as a function of their
$\beta_i$ obtained by regressions with respect to the excess return to
the effective S\&P323 index. Under the CAPM hypothesis, one should
obtain a straight line with slope $\E \[r_{SP323} - r_0\]$ ($-13.1\%$
per month) and zero additive coefficient at the origin. The straight
line is the regression $y = 0.88\% -13.5\% \cdot x$.
A standard statistical test shows that the value $0.88\%$ of the intercept
at the origin is not statistically significant from zero. Together
with the reasonable agreement between the slope of the regression and the excess
expected returns of the S\&P323 index, this would give a positive
score for the CAPM. This is perhaps surprising considering the
biases distribution of alpha's shown in figure \ref{InterceptRealData_ExcessReturnSP323}.
This suggests that this standard expected return/beta tests examplified in figure 
\ref{InterceptRealData_ExcessReturnSP323} has not large power.

\begin{figure}
\includegraphics[width=14cm]{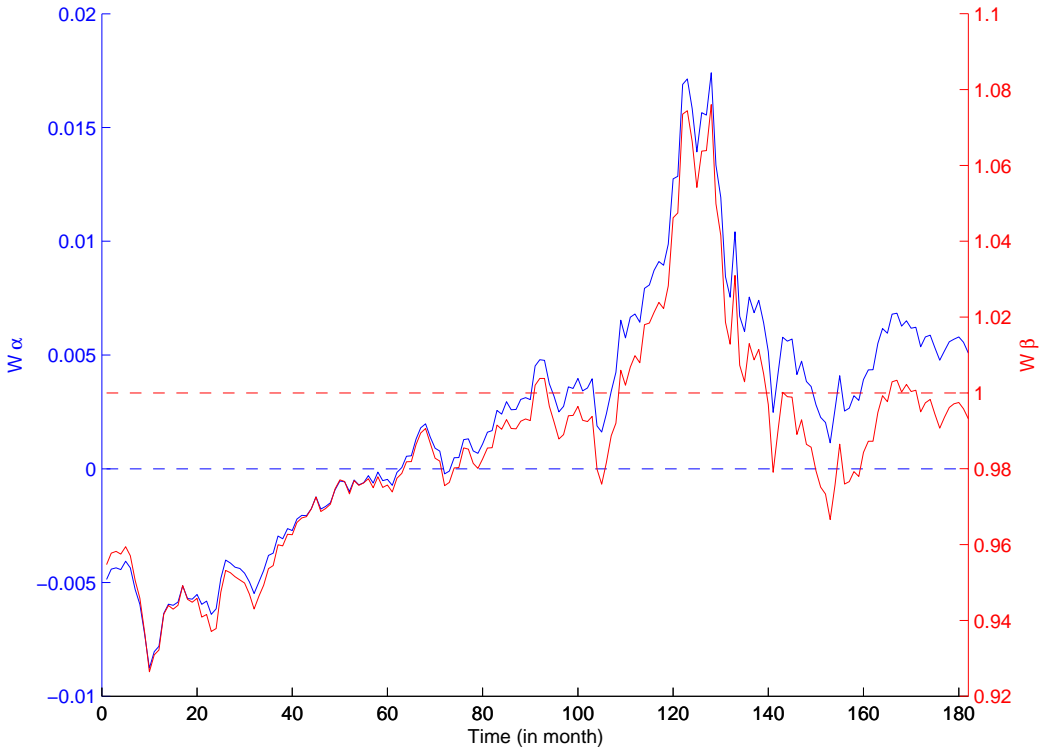}
\caption{Time evolution of $\tilde w_t' \vec{\tilde \beta}$ (red and right vertical scale) and  
$\tilde w_t' \vec{\tilde \alpha}$ (blue and left vertical scale) over the period 
from January 1990 to February 2005 which includes 182 monthly values.
$\tilde w_t$ is the vector of weights of the 323 stocks in our
effective S\&P323 index which evolves at each time step according
to the capitation of each stock. $\vec{\tilde \beta}$ and 
$\vec{\tilde \alpha}$ are the two vectors of beta's and alpha's obtained
from the regressions used in figures \protect\ref{InterceptRealData_ExcessReturnSP323}
and \protect\ref{CapmRealData_ExcessReturnSP323}.
According to the self-consistency conditions (\ref{eq:msfjilj}) and 
(\ref{mmkvkmlala}), the dynamical consistency of the CAPM should lead
to $\tilde w_t' \vec{\tilde \beta} = 1$  and 
$\tilde w_t' \vec{\tilde \alpha}=0$ at all time periods.}
\label{Constraints}
\end{figure}

As a complement, one can use the self-consistency conditions 
$\tilde w_t' \vec{\tilde \beta} = 1$ (expression \ref{eq:msfjilj}) and 
$\tilde w_t' \vec{\tilde \alpha}=0$ (expression \ref{mmkvkmlala})
to perform empirical tests. As explained in 
section \ref{sec:capm}, the dynamical consistency of the CAPM
imposes that these two relationships should hold at each time step
for the proxy of the market portfolio. We have thus calculated
$\tilde w_t' \vec{\tilde \beta}$ and  $\tilde w_t' \vec{\tilde \alpha}$,
where $\tilde w_t$ is the vector of weights of the 323 stocks in our
effective S\&P323 index which evolves at each time step according
to the capitation of each stock while $\vec{\tilde \beta}$ and 
$\vec{\tilde \alpha}$ are the two vectors of beta's and alpha's obtained
from the regressions used in figures \ref{InterceptRealData_ExcessReturnSP323}
and \ref{CapmRealData_ExcessReturnSP323}. Figure \ref{Constraints}
shows the time evolution of $\tilde w_t' \vec{\tilde \beta}$ and  
$\tilde w_t' \vec{\tilde \alpha}$ over the period 
from January 1990 to February 2005 which includes 182 monthly values.
The deviations respectively from $1$ and $0$ are significant, as
shown by a standard Fisher test. 
The close connection between the time varying average alpha and beta
shown in Figure \ref{Constraints} results from their common 
dynamics through the evolution of the weights $\vec w$.

The variable $\tilde w_t' \vec{\tilde \beta}$ can be interpreted
as the average beta of the stocks in the self-consistent market proxy.
A value different from $1$ suggests that the market is out of equilibrium.
In particular, if $\tilde w_t' \vec{\tilde \beta} > 1$, this can
be interpreted as an ``over-heating'' of the market with 
the existence of positive feedback. Interestingly,
this occurs just about two years before the peak of the Internet
bubble in April 2000. It then took about two years after the peak
to recover an equilibrium. Since early 2003, the market seems
to have remained approximately at equilibrium according to this metric.

\subsection{Tests on a synthetically generated market}

\begin{figure}
\includegraphics[width=14cm]{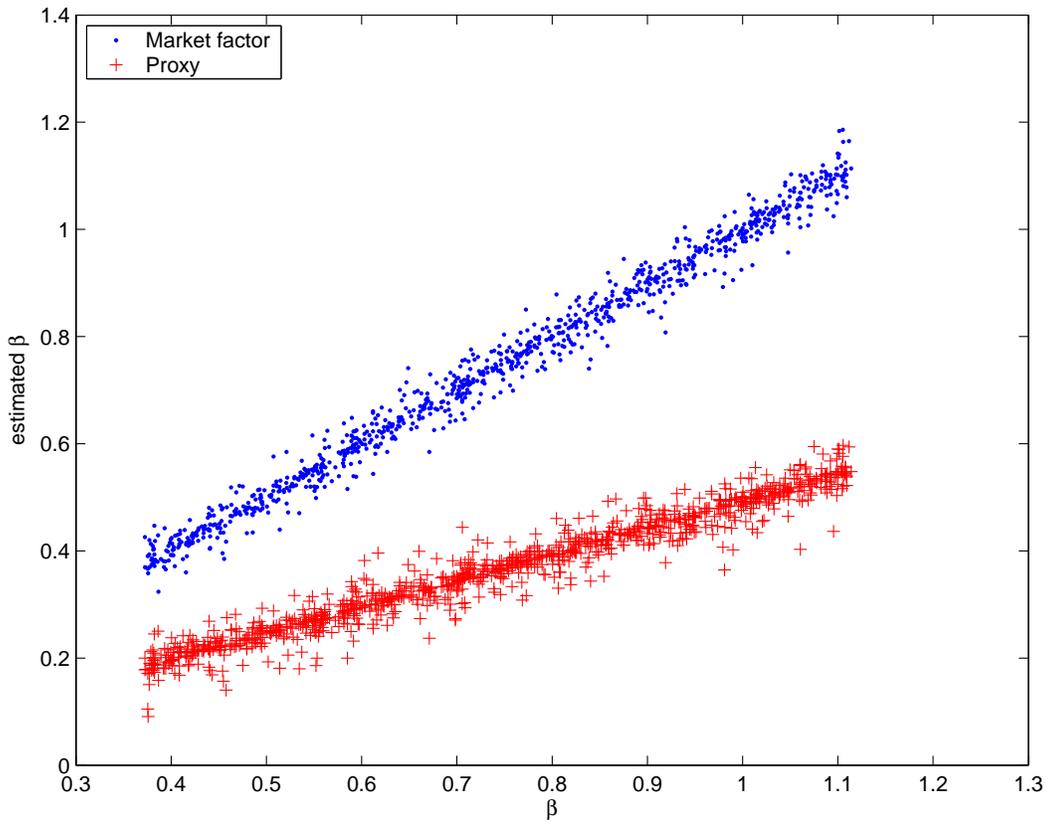}
\caption{Synthetic tests on an artificial
market of 1000 synthetic assets with properties adjusted
to mimic those of the real US market. 
The plot shows the estimated beta's obtained from the regression of the asset
returns on the returns of the market portfolio (blue dots) and on the
returns of the proxy (red crosses), as a function of the true beta's.
The upper straight line corresponds to the ideal case where the estimated
beta's equal the true beta's. The lower straight line is
the predicted dependence (\ref{eq:dhml}) of the beta's estimated with the 
proxy as a function of the true beta.}
\label{figproxy1}
\end{figure}

In order to investigate the sensitivity of these tests, and in particular
the impact of using a proxy for the market portfolio, 
we have constructed a toy (synthetic) market in which 1000 assets are traded and such
that their returns at time $t$ obey equation (\ref{eq:eriugh2}) with
the constraints (\ref{mvmlsl}). The weight of each asset in the
market portfolio is drawn from a power law with tail index equal to one,
in accordance with empirical observations on the distribution of firm
sizes (Axtell, 2001), and then renormalized so that the weights sum up to
one. For the purpose of illustration and easiness in testing, we
impose that the composition of the market remain constant, i.e., the
economy is stationary. The interest in this condition is that we can
then study the pure impact of not observing the true market but only
the proxy constructed on a subset of the whole universe of assets.
The daily return on the synthetic market factor follows a
Gaussian law with mean and standard deviation equal to the mean and the
standard deviation of the daily return on the S\&P500 over the time
period from July 1962 to December 2000, namely $0.037\%$ and $0.90\%$
respectively. The $\beta$'s are also randomly drawn from a uniform law
with mean equals to one and are such that they satisfy the normalization
condition (\ref{mvmlsl}). It can be seen in figure~\ref{figproxy1} that
the $\beta$'s range between $0.35$ and $1.15$, which is reasonable if we
refer to the values usually reported in the literature. Finally, the
residuals $\vec \epsilon_t$ are drawn from a degenerate multivariate
Gaussian distribution ({\it i.e.}, the rank of its covariance matrix is
$N-1=999$), so that they fulfill the orthogonality condition
(\ref{mvmlsl}). The variances and covariances of these residuals have
been fixed in such a way that they are of the same order of magnitude as
the variances and covariances of the residuals estimated by linear
regression of our basket of 25 assets on the S\&P500. Thus, the values
given by our toy market are expected to be consistent with the values
observed on the actual market if the description by a one factor model
has some merit.

Using the OLS estimator, we have first performed a regression
with respect to the true market portfolio, whose
composition is assumed to remain constant as we said. Then, we
have constructed an arbitrary portfolio and have considered it to
be the proxy of the market portfolio. We have then performed
the linear regression of the assets returns on the proxy returns. Figure
\ref{figproxy1} compares the estimated beta's obtained
from the regression of the asset returns on the returns of the market
portfolio with those obtained from the regression
on the returns of the proxy, as a function of the true beta's.
The regression on the market factor gives a line with unit slope and
zero intercept, as expected from the construction of the synthetic market.
The regression on the proxy returns gives also a straight line, as
predicted from the linear relation between $\vec \beta$ and $\vec{\tilde
\beta}$ given by (\ref{eq:dhml}) in Appendix C. Figure \ref{figproxy1}
provides a verification of the properties put by construction in our
synthetic market. Obviously, no one would be able
to perform this verification on real data since the market portfolio
and thus the true beta's are unknowable.

\begin{figure}
\includegraphics[width=14cm]{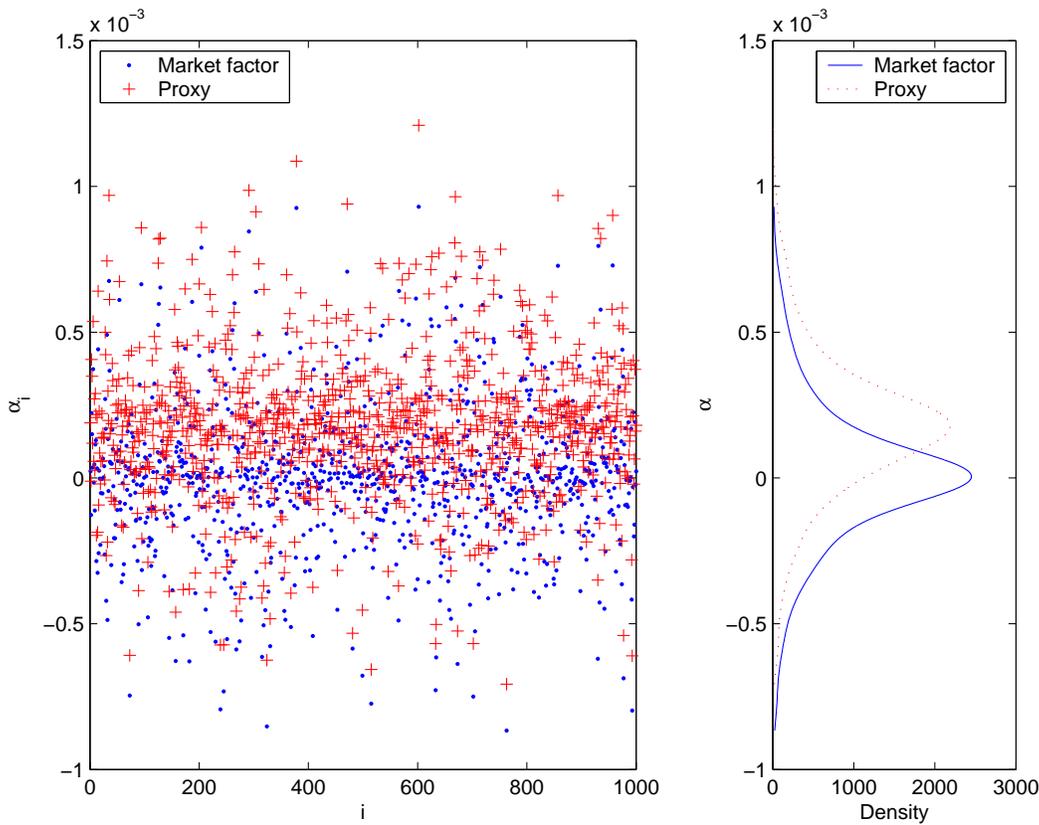}
\caption{Synthetic tests on an artificial
market of 1000 synthetic assets with properties adjusted
to mimic those of the real US market. Left panel: Population
of the intercepts of the regression of expected stock returns
versus the market return (blue dots) or versus the proxy return (red crosses)
in our synthetic market. The abscissa is an arbitrary indexing
of the 1000 assets of our artificial market. The estimated probability density functions of
the two population of alpha's are shown on the right panel
and illustrate the existence of a systematic bias for the proxy's alpha's.}
\label{figproxy2}
\end{figure}

Figure \ref{figproxy2} shows the population
of the intercepts of the regression of expected stock returns
versus the market return or versus the proxy return in our synthetic market. These
intercepts are presented  
as a function of the (arbitrary) indices of the 1000 assets. For the regression
on the market factor, one can observe as expected a scatter around zero.
For the regression on the market proxy,
the intercepts are, on average, all significantly different from zero.
As expected, the orthogonality and normalization conditions
$\tilde w' \vec \alpha=0$  and $\tilde w' \vec {\tilde \beta}=1$
are satisfied, providing a verification of the validity of the numerical implementation
of the model for these synthetically generated data. Thus, figure \ref{figproxy2}
confirms that a universe of assets which by construction obeys the CAPM exhibits
non-zero alpha intercepts (which take apparently random values) when using an arbitrary proxy.
This result can be compared with the empirical analog shown in figure
\ref{InterceptRealData_ExcessReturnSP323}.

\begin{figure}
\includegraphics[width=14cm]{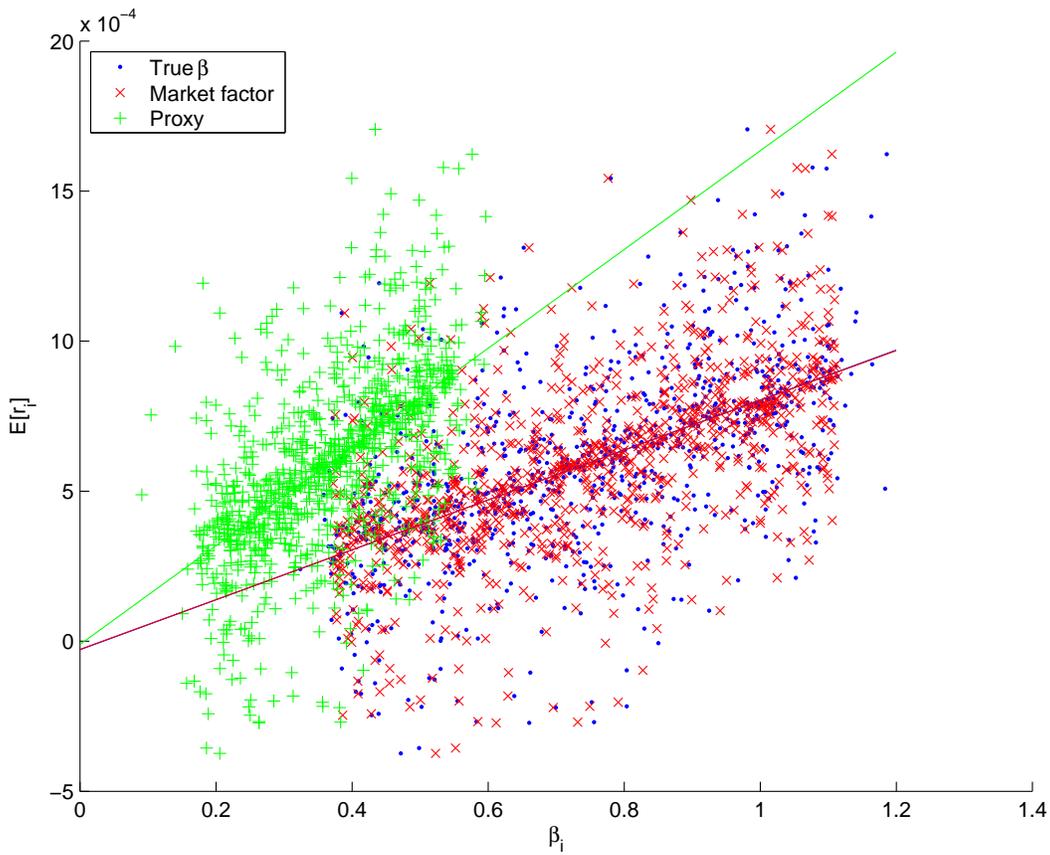}
\caption{Synthetic tests on an artificial
market of 1000 synthetic assets with properties adjusted
to mimic those of the real US market. Individual expected returns $\E \[r_i\]$
for each of the 1000 assets (i) as a function of the
true $\beta_i$'s (blue dots), (ii) as a function of the $\beta_i$'s obtained by 
regression on the true market (red crosses $\times$) and by regression on the proxy 
(green $+$). The straight lines are the linear regressions.}
\label{figproxy3}
\end{figure}

Figure \ref{figproxy3} shows the individual expected returns $\E \[r_i\]$
for each of the 1000 assets (i) as a function of the
true $\beta_i$'s, (ii) as a function of the $\beta_i$'s obtained by 
regression on the
true market and (iii) by regression on the proxy. As expected, the dependence
of the expected returns on the true beta's and on the beta's obtained from the
true market portfolio follows the CAPM prediction, but with rather significant
fluctuations. The scatter of the dependence of the expected returns on the
beta's determined from the proxy is larger but one can still observe
a well-defined linear dependence with a zero intercept,
and a slope different from the
expected return $\E\[\tilde r_t\]$ of the portfolio proxy,
as predicted in expression (\ref{mbnsjkswls}). This seems to 
justify why the bias in the distribution of alpha's does not seem to
affect the existence of the standard expected return/beta test shown in figure 
\ref{InterceptRealData_ExcessReturnSP323}.

\subsection{On the orthogonality and normality conditions}

To summarize, the condition of self-consistency leads to the
orthogonality and normality conditions (\ref{mvmlsl}) for the
mono-factor model and to (\ref{eq:sdfkuh},\ref{eq:pgho}) for the
multifactor model when the market portfolio is known. 
The orthogonality and normality conditions still hold when only a
market proxy is available and they take the form (\ref{eq:msfjilj})
together with the additional orthogonality constraint
(\ref{mmkvkmlala}). This suggests to use the orthogonality and normality
conditions as new tests of the CAPM in the real-life situation where the
market portfolio is not known and a somewhat arbitrary proxy is used.
The motivation of these tests stems from the fact that they are not
affected by the problem of using a proxy which is different from the real market
factor, in contrast with the problem on the standard
test of the CAPM made explicit in figure
\ref{figproxy3}.
Concretely, this suggests to complement
the standard expected excess return versus beta, by
tests checking the validity of the orthogonality and normality
conditions when using for the proxy, {\em not} the S\&P500, but {\em
any} portfolio constructed on the assets used in the test. A
test of the CAPM would then consist in testing the normalization and
orthogonality conditions (\ref{eq:msfjilj}-\ref{mmkvkmlala}), which
should hold for any such proxy portfolio.

It turns out however that the OLS estimated
intercepts $\hat {\vec \alpha}$, the estimated $\beta$'s $\hat {\vec \beta}$ 
and the estimated residuals $\hat {\vec \eta}$ 
of a basket of assets necessarily satisfy the constraints
(\ref{eq:msfjilj}-\ref{mmkvkmlala}) when the proxy used as the regressor
is a portfolio build on these same assets. Let us denote by $Y$ the matrix
which stacks the returns of the basket of the $N$ assets under consideration, by
$X$ the matrix of the regressors, by $B$ the matrix of the regression
coefficients and by $U$ the matrix which stacks the vectors of the
residuals:
\be
Y=
\(
\begin{array}{c}
{\vec r_1}'\\
\vdots\\
{\vec r_T}'
\end{array}
\),\quad
X=
\(
\begin{array}{cc}
1 & r_m(1)\\
\vdots & \vdots\\
1 & r_m(T)
\end{array}
\),\quad
B=
\(
\begin{array}{ccc}
\alpha_1 & \cdots & \alpha_N\\
\beta 1 & \cdots & \beta_N
\end{array}
\),\quad
U=
\(
\begin{array}{c}
{\vec \eta_1}'\\
\vdots\\
{\vec \eta_T}'
\end{array}
\)~,
\ee
so that, if $\vec r_m$ denotes the vector of the returns on
any portfolio $W$ made of our $N$ assets only, we have
\be
\(
\begin{array}{c}
r_m(1)\\
\vdots\\
r_m(T)
\end{array}
\) = Y W~.
\ee
With these notations, the  linear regression equation reads $Y = XB + U$. The OLS
estimators of $B$ and of $U$ are then respectively
\be
\hat B = \(X^t X\)^{-1} X^t Y
\ee
and
\be
\hat U = Y - X \hat B = \[ {\rm Id} -X \(X^t X\)^{-1} X^t \]Y,
\ee
It is then easy to show that
\be
\hat B W = \( \begin{array}{c} 0\\1 \end{array}\), \quad {\rm et} \quad \hat U W = \vec 0~,
\ee
which are nothing but the constraints
(\ref{eq:msfjilj}-\ref{mmkvkmlala}) in matrix form. Their derivation
involves the same kind of algebraic manipulations as those employed in
Appendix E in the next section and are thus not repeated here.
Therefore, given any
portfolio made of the subset of assets under consideration only, the OLS
estimator automatically provides estimates which fulfill the
self-consistency constraints. This prevents us from using these constraints as
a way to test the CAPM. However, this derivation shows that, assuming that the CAPM holds,
the OLS method provides a simple way to self-consistently assess
the parameters of the model by using proxy portfolios made only of the
assets which are used in the CAPM regressions.

\section{Multi-factor models  \label{mnvgkls}}

\subsection{Orthogonality and normality conditions}

Extending section \ref{mmslsl;}, we now
investigate the implications of using portfolio proxies
for the explanatory factors in the multi-factor model analyzed in
section \ref{mgmsa;a}.

Let us first assume that the individual asset
returns can be explained by {\em exactly} $q$ factors. Then, $q$ factor proxies
are built by defining $q$ portfolios of the traded assets. Let us
denote by $\tilde W$ the matrix whose columns represent the $q$
portfolios and by $\vec v_t$ the vector of the $q$ proxies.
Appendix D shows that, similarly to the result (\ref{mgmlss}) obtained
for the one-factor model, a non-zero intercept $\vec \alpha$ appears
in the regression of the vector of asset returns with respect to
the $q$ proxies in the vector $\vec v_t$ (see expression (\ref{eq:pdfig})).
In addition, the normalization condition
\be
\label{eq:pdgho2}
\tilde W' \tilde B = {\rm Id}
\ee
and the two orthogonality conditions
\be
\label{eq:pdgho3}
\tilde W' \vec \alpha = \vec 0, \quad {\rm and} \quad \tilde W' \vec 
\nu_t=\vec 0,
\ee
hold, where $\nu_t$ is the vector of the residuals of the multivariate
regression on the vector of the $q$ proxies $v_t$.

A priori, we do not know how many factors are needed but there are 
standard tests in factor analysis that provide some estimates
of the number of factors (Connor and Korajzcyk, 1993; 
Bai and Ng, 2002). It is possible to encounter 
a situation where the number $r$ of portfolio proxies is
different from the true number $q$ of factors. The case $r<q$ corresponds
to market incompleteness. Let us discuss the situation where $r>q$.
In this case, equations (\ref{eq:pdgho2}--\ref{eq:pdgho3}) still hold,
as shown in Appendix E, but a difficulty arises from the fact that the
matrix  $\tilde W' B$ is not a $q\times q$ matrix anymore, it is a $r
\times q$ matrix, where
$r>q$ is the number of chosen factor proxies. As a consequence,
$\(\tilde W' B\)^{-1}$ does not exist and has to be replaced by its
(left) pseudo-inverse. As previously, a non-zero intercept $\vec 
\alpha$ also appears
in the regression of the vector of asset returns with respect to
the $q$ proxies. The orthogonality and normalization conditions still hold,
as shown in Appendix E.

\subsection{Self-consistent calibration of the multi-factor model 
and principal component analysis (PCA) \label{nvbkdd}} 

Let us assume the existence of $Q$ factors which can be replicated by
$Q$ portfolios $W_i$ (the market is complete). Let $W$ be the matrix
which stacks all these
portfolios: $W=\(W_1, W_2, \ldots, W_Q\)$. We again denote $\vec r_t$
as the vector of
excess returns of the $n$ assets over the risk free rate\footnote{
If for instance the APT is true (i.e., there are no arbitrages available), then one
does not need to subtract means for the intercept in (\ref{ngjhk;f}) to be zero.}, 
$\vec u_t =W' \vec r_t$
is the set of factors and $B$ is the matrix of beta's. This
defines the model (\ref{eq:kukggf}):
\bea
\vec r_t &=& B \vec u_t + \vec \epsilon_t~,\\
&=& B W' \vec r_t + \epsilon_t~,
\label{ngjhk;f}
\eea
where the intercept is set to zero, which is always possible provided
that we subtract the mean value of $\vec r_t$.
Appendix F shows how to estimate the beta's $B$ and the $Q$ replicating
portfolios $W=\(W_1, W_2, \ldots, W_Q\)$ by using the properties
\bea
W' \vec 1_N &=& \vec 1_Q~,    \label{mvmb}\\
W' B &=& {\rm Id}_Q~,     \label{nbjklls;} \\
W' \vec \epsilon_t &=& 0~.  \label{mgmlsww}
\eea
The first property (\ref{mvmb}) just expresses the normalization of
the portfolio weights.
The two other properties are the normalization and orthogonality 
conditions derived from
the self-consistency condition that the factors can be replicated
by portfolios constituted of the assets that they are supposed to explain
(see (\ref{mvmlsl}) for the one-factor case and
(\ref{eq:sdfkuh},\ref{eq:pgho}) for the multi-factor case).

Appendix F first derives the relation (\ref{wpoqpuvqa;nqf})
\be
W= B \[B'B\]^{-1}~,   \label{qsiquqiaa2}
\ee
between the matrix $W$ of weights and the matrix $B$ of beta's,
showing the dependence between $W$ and $B$ resulting from the
self-consistency conditions.
Finally, $B$ and $W$ can be constructed as (\ref{mnbkl},\ref{bnjewla})
\bea
B &=& P' U \nabla V'~,   \label{mnbkl2}\\
W &=& P' U \nabla^{-1} V'~.   \label{bnjewla3}
\eea
The matrix $P$ is specified by the decomposition $RR'=P'DP~$
given in (\ref{mbm,;w;w}),
where $R= \(\vec r_1, \vec r_2, \ldots, \vec r_T \)$ is a $N \times T$ matrix
and $D$ is the diagonal matrix with elements equal to the eigenvalues
of $RR'$. The matrix $U$ is also fixed by (\ref{m,h,a;,s}), i.e., it has
its first $Q$ upper diagonal elements equal to $1$ and all its other elements
equal to zero. The matrix $V$ is not uniquely fixed, reflecting in this way
the rotational degeneracy of the $Q$ factors. Indeed, matrix $V$ can be
any $Q \times Q$ orthogonal matrix whose lines add up to a
non vanishing constant.

Expression (\ref{ngjhk;f}) with (\ref{mnbkl2},\ref{bnjewla3}) offers a
practical decomposition of the market risks, using a multi-factor model
generalizing the CAPM. It is useful to compare it with other available
methods. It is customary in the financial literature to distinguish
between model-driven and data-driven constructions of risk factors
(Loretan, 1997). The CAPM is a good example of a model-driven method
which imposes strict relationship between asset prices. On the other
hand, the Principal Components Analysis (PCA) method is the archetype of
data-driven methods, which enjoys widespread use among statistical
practitioners (Dunteman, 1989; Jolliffe, 2002). PCA is frequently
employed to reduce the data dimensionality to a tractable value
without needing strong hypotheses about the
nature of the data generating process. Now, the reader familiar
with PCA will notice that expression (\ref{ngjhk;f}) with (\ref{mnbkl2},\ref{bnjewla3})
provides a decomposition of risk components which {\it is nothing but} 
the decomposition obtained by using PCA! In other words, 
this section together with Appendix F has shown that
a multi-factor
analysis implemented {\em with the self-consistency condition} is
equivalent to  the empirical methodology of analyzing baskets of assets using
PCA. 

In general, there are no any necessary connection between 
data-driven and model-driven constructions of risk factors. But, 
as soon as one uses a factor
model, if the factors can be indeed expressed in terms of the assets
themselves they are supposed to explain (as in the Fama/French 3-factor
model) which is nothing but the self-consistency condition, then it
follows automatically and necessarily that there is a connection between
the factor model and the PCA: in fact, 
the factor analysis and the PCA are one and the same. This shows
again the strong constraint that the self-consistency condition provides.
This provides a direct link between model-driven and data-driven
constructions of risk factors: one of the best
representative of model-driven risk factor decomposition methods (the
multi-factor model with self-consistency) is one and the same 
as one of the best examples of data-driven risk factor decomposition methods
(the PCA). This correspondence implies that PCA will therefore suffer
from the same limitations as the CAPM and its multi-factor generalization,
namely lack of out-of-sample explanatory power and predictability.
The exact correspondence between self-consistent multi-factor models
and PCA justifies claims on the empirical and practitioner literature\footnote{
see for instance 
\url{http://www.perfectdownloads.com/business-finance/investment-tools/pickstock.htm}
and \url{http://www.apt.com/en/aboutus/theaptapproach.html}}
that PCA may be an implementation of the arbitrage pricing theory (APT)
(Ross, 1976; Roll and Ross, 1984; Roll, 1994).
Our result also suggests that using PCA to pre-filter the data before
a factor decomposition is misconceived since both PCA and factor decomposition
are one and the same thing. It might however be useful 
in nonlinear factor decomposition, as suggested from 
previous nonlinear dynamic studies
(Broomhead and King, 1986; Vautard et al., 1992; Chan and Tong, 2001).

PCA is theoretically better in one sense: it works with the raw
covariance matrix of returns and hence should uncover any factors
present in that matrix.  The same cannot be said about approaches 
in terms of a fixed pre-determined number of factors. It is quite possible
that the later approaches will fail to uncover important
factors.  However, PCA has a disadvantage because it is difficult to
estimate when allowing for time variation in the true covariance matrix.
This is in that sense that the factor models are more tractable.

\section{Discussion and conclusion}

We have structured the presentation of factor models in the light
of the self-consistency condition. Starting from arbitrary factor models, 
internal consistency
requirements have been shown to impose strong constraints on the coefficients of the
factor models. These requirements merely express the fact that the
factors employed to explain the changes in assets prices are themselves
combinations of these securities. These conditions read
\be
\label{eq:skmjkh}
W_t'B_t = {\rm Id}, \quad {\rm and} \quad W_t' \vec \epsilon_t = 0.
\ee
In addition, when  proxies of the market factors are used instead of the
factors themselves, a non-vanishing
intercept $\vec \alpha$ appears which satisfies the third constraint
\be
\label{eq:dgpoti}
W' \vec \alpha =0.
\ee
These constraints are appealing and it would have been natural to use
them to test the adequacy of the factor-models. However, 
they are automatically fulfilled by the regression (i) on a proxy
which is a portfolio whose composition is constant through time and is
restricted to the subset of assets under consideration and (ii) on the
factors derived from the PCA, when one uses this statistical method to
select the relevant explaining factors. Thus, on the one end, these
constraints do not allow to test the CAPM (or the multi-factor models),
which remains untestable unless the entire market is considered, as
first stressed by Roll (1977); nevertheless, on the other hand, the OLS
estimator and the PCA provides a consistent method to assess the value of
the different parameters of the problem.

Now, to escape from this self-referential approach which consists in
regressing the assets returns on the returns on a portfolio made of the
assets under consideration with constant proportion, one has to use a
proxy with non-constant composition, such as the Standard \& Poor's 500 index.
In such a case, the normalization and orthogonality conditions
(\ref{eq:skmjkh}-\ref{eq:dgpoti}) must hold at each time $t$. Thus, for
a number of periods $t$ larger than the number $N$ of assets
constituting the proxy, the number of constraints is larger than the
number of parameters $\alpha_i$'s and $\beta_i$'s to estimate. This
implies that $\vec \beta$ and $\vec \alpha$ can not be constant, unless
the time varying vectors of market weights $\vec w_t$ ``live'' in a
subspace of ${\mathbb R}^{N}$ which is orthogonal to $\vec \alpha$ and
such that $\vec w_t' \cdot \vec \beta =1$ (given by (\ref{mvmlsl}) for
the mono-factor model, by (\ref{eq:sdfkuh},\ref{eq:pgho}) for the
multifactor model when the market portfolio is known and by
(\ref{eq:msfjilj},\ref{mmkvkmlala}) when only a market proxy is
available).

This condition raises questions on the dynamic consistency of the
CAPM. As stressed, and then immediately swept under the carpet, at the
end of section~\ref{sec:capm}, the equilibrium imposes a dynamic
constraint on the composition of the market portfolio: on the one hand,
it is endogenously determined by the investors' anticipations according
to formula (\ref{eq:dghpo}) while, on the other hand, the market
portfolio must be related to the market capitalization of each asset,
which reflects the economic performance of the industry. Thus, the
relation (\ref{eq:dfjh}) must hold. It can be rewritten as
\be
\label{eq;sdhg}
w_{t+1}^i =  w_t^i \cdot \frac{1 + r_0 + \beta_i r_m(t) + u_t^i}{1+ r_0 + r_m(t)}.
\ee
This relation would be compatible with the normalization
condition at times $t$ and $t+1$ if and only if $\sum_{i=1}^N \beta_i
w_t^i = \sum_{i=1}^N \beta_i w_{t+1}^i=1$ which would imply that
\be
r_m(t) = \frac{\sum_{i=1}^N \beta_i w_t^i u_t^i}{1 - \sum_{i=1}^N w_t^i \beta_i^2}.
\label{ngsla;l}
\ee
But now, what could justify such a relation between the market return
and the residuals. They have been assumed independent (or at least
uncorrelated) up to now. Recall that our basic assumption was that
$r_m(t)$ is exogenously fixed by the economic environment.

In this respect, it seems imperative to give up the assumption of a
constant $\vec \beta$. But, as a consequence, it becomes necessary to
specify a dynamics for $\vec \beta_t$. Several works have started
addressing this question (Blume, 1971; 1975; Ohlson and
Rosenberg, 1982; Lee and Chen, 1982; Bos and Newbold, 1984; Simmonds et al.,1986;
Collins et al., 1987) and have proved the merit of this
approach. With regard to this question, both eq. (\ref{eq:threj2}) and
figures~\ref{XOM} and~\ref{Renormalized} suggest the existence of a well
defined average $\beta$. Besides, considering that the volatility of the
assets returns is mean-reverting, which is a well-known stylized fact
(Satchell and Knight, 2002; Figlewski, 2004), eq. (\ref{eq:threj2}) 
shows that such an assumption should
also hold for the dynamic of $\beta_t$ \footnote{To get this result, 
let us start from expression (\ref{eq:threj2}) for the vector $\vec \beta$.
Let us assume that the matrix
$\Omega$ has a dynamics of its own which is mean-reverting,
$\Omega(t) = \Omega_0 + f(t) O$, 
where we assume that the time dependence is in the scalar factor $f(t)$,
while $O$ is a constant matrix. Let us assume that $f(t)$ is small, so that
$f(t) O$ constitutes a perturbation to $\Omega_0$. Expression
(\ref{eq:threj2}) can be expanded to first order in powers of $f(t)$ to obtain
$\beta(t) = C (1+f(t) {\bar O}) + \beta_0$, 
where $C$ and ${\bar O}$ are constant matrices which can be expressed in terms of 
$O, \Omega_0, {\vec \alpha}$ and ${\vec \beta}_0$. This shows that,
if $f(t)$ is mean-reverting, then $\beta(t)$ is also
mean-reverting.} 

Finally, the normalization condition shows that $\vec \beta_t$ can be
written as the sum of two terms
\be
\vec \beta_t = \frac{\vec w_t}{||\vec w_t ||^2} + \vec \beta_t^\dagger,
\ee
where $\vec w_t' \cdot \vec \beta_t^\dagger =0$. The first term, $\vec w_t/||\vec w_t
||^{2}$, is directly related to the Herfindahl index, i.e. the
concentration, of the market portfolio. So, everything else taken
equal, the risk premium increases when the level of diversification of
the market decreases. As a first approximation, $\vec \beta_t^\dagger$
could be taken constant, so that the dynamics of $\vec \beta_t$ could be
easily related to the dynamics of the market portfolio, which is a
predictable quantity ($\vec w_t$ is known at time $t-1$, by use of
(\ref{eq;sdhg})).

As mentioned briefly in the introduction, there is another interesting
consequence of the self-consistency condition when an addition ingredient
holds, namely when the distribution of the capitalization of firms is
sufficiently heavy-tailed. In such case which seems to be relevant
to real economies, assuming that 
a general complete equilibrium with no-arbitrage holds,  
then one finds that arbitrage-pricing is actually
fundamentally inconsistent with equilibrium even for arbitrary large
real economies: there exists a significant non-diversifiable risk 
which is however not priced by the market (Malevergne and Sornette, 2006b).
This result is based on 
the self-consistency condition discussed at length in this paper, which
leads mechanically to
correlations between return residuals which are equivalent to 
the existence of a new ``self-consistency'' factor. Then, 
when the distribution of the capitalization of firms is
sufficiently heavy-tailed, it is possible to show, using methods
associated with the generalized central limit theorem, that the 
``self-consistency''
factor does not disappear even for infinite economies and may produce
significant non-diversified non-priced risks for arbitrary
well-diversified portfolios. 
For economies in which the return residuals are function of the
capitalization of firms, the new self-consistency factor provides
a rationalization of the SMB (Small Minus Big) factor introduced by
Fama and French.


\pagebreak

\section*{Appendix A: derivation of the CAPM relations
(\ref{eq:pjzer2}) and (\ref{eq:threj2})}

Let us consider the single factor model (\ref{eq:capm}) together with the
self-consistency condition (\ref{mvgmlsl}) that the market is constructed
over the observable universe of securities. Then,
left-multiplying (\ref{eq:capm}) by the market portfolio $\vec w_t'$ yields
\be
\label{eq:market}
r_m(t) =\frac{\vec w_t' \cdot \( \vec \alpha + \vec \epsilon_t\)}{1 -
\vec w_t' \cdot
\vec \beta^0}, \quad {\rm and} \quad \E \[r_m(t) \] =
\frac{\vec w_t' \cdot \vec \alpha }{1 - \vec w_t' \cdot
\vec \beta^0}.
\ee
By substitution in (\ref{eq:capm}), we obtain
\be
\label{eq:return}
\vec r_t = \( {\rm Id} + \frac{\vec \beta^0 \cdot \vec w_t' }{1 - \vec
w_t' \cdot \vec \beta^0} \) \( \vec \alpha + \vec \epsilon_t\), \quad
{\rm and} \quad
\E \[\vec r_t \] = \( {\rm Id} + \frac{\vec \beta^0 \cdot \vec w_t' }{1 - \vec
w_t' \cdot \vec \beta^0} \) \vec \alpha.
\ee

Assuming that the investors aim at achieving the dynamic 
mean-variance program $({\cal P})$,
Li and Ng (2000) have shown that they all invest a part of their 
wealth in the risk-free
asset and the remaining in a portfolio made of risky assets only, 
whose composition is given by
\be
\vec w_t = \frac{\Sigma_t ^{-1} \E \[\vec r_{t} \]}{\vec 1' \Sigma_t
^{-1}\E \[\vec r_{t} \]} =
\frac{\Sigma_t ^{-1}\( {\rm Id} + \frac{\vec \beta^0 \cdot \vec w_t' }{1 - \vec
w_t' \cdot \vec \beta^0} \) \vec \alpha}{\vec 1' \Sigma_t ^{-1}\( {\rm Id} +
\frac{\vec \beta^0 \cdot \vec w_t' }{1 - \vec
w_t' \cdot \vec \beta^0} \) \vec \alpha},
\label{mgmllr}
\ee
where $\Sigma_t$ is the conditional covariance matrix of the returns
$\vec r(t)$. Now, if an equilibrium is reached at every time $t$, $\vec
w_t$ represents the market portfolio at this time.

  From  (\ref{eq:return}), one easily obtains that, conditional on the
observations up to time $t-1$,
\be
\label{eq:fsjkg}
\Sigma_t = \( {\rm Id} + \frac{\vec \beta^0 \cdot \vec w_t' }{1 - \vec
w_t' \cdot \vec \beta^0} \) \Omega_t \( {\rm Id} + \frac{\vec w_t  \cdot
\vec {\beta^0}'}{1 - \vec w_t' \cdot \vec \beta^0} \).
\ee
By use of Shermann-Morrison inversion formula
(see Golub and Van Loan 1996, for instance), we have
\be
\label{eq7}
\( {\rm Id} + \frac{\vec \beta^0 \cdot \vec w_t' }{1 - \vec w_t'
\cdot \vec \beta^0} \)^{-1} =
      {\rm Id} - \vec \beta^0 \cdot \vec w_t',
\ee
so that
\be
\label{eq8}
\Sigma_t^{-1} = \( {\rm Id} - \vec w_t \cdot \vec {\beta^0}'\)
\Omega_t^{-1} \({\rm Id} - \vec \beta^0 \cdot \vec w_t' \).
\ee
Substituting this expression in equation (\ref{mgmllr}) yields
\bea
\vec w_t &=& A^{-1} \cdot \( {\rm Id} - \vec w_t \cdot \vec {\beta^0}'\)
\Omega_t^{-1} \({\rm Id} - \vec \beta^0 \cdot \vec w_t' \) \( {\rm Id} +
\frac{\vec \beta^0 \cdot \vec w_t' }{1 - \vec
w_t' \cdot \vec \beta^0} \) \vec \alpha\\
&=& A^{-1} \cdot \( {\rm Id} - \vec w_t \cdot \vec {\beta^0}'\)
\Omega_t^{-1} \vec \alpha~,
\label{eq:trz}
\eea
where $A$ is a scalar equal to
\bea
A &=& \vec 1'
\( {\rm Id} - \vec w_t \cdot \vec {\beta^0}'\)
\Omega_t^{-1} \({\rm Id} - \vec \beta^0 \cdot \vec w_t' \) \( {\rm Id} +
\frac{\vec \beta^0 \cdot \vec w_t' }{1 - \vec
w_t' \cdot \vec \beta^0} \) \vec \alpha\\
&=& \vec 1'
\( {\rm Id} - \vec w_t \cdot \vec {\beta^0}'\)
\Omega_t^{-1} \vec \alpha~.
\label{mgele}
\eea
The two equalities (\ref{eq:trz}) and (\ref{mgele}) result
from (\ref{eq7}).

Expanding the right hand side of (\ref{eq:trz}), we obtain
\be
\vec w_t = A^{-1} \Omega_t^{-1} \vec \alpha - A^{-1} \( \vec
{\beta^0}' \Omega_t^{-1}
\vec \alpha \) \vec w_t
\ee
so that
\be
\vec w_t = \frac{1}{A + \vec {\beta^0}' \Omega_t^{-1} \vec \alpha}~
\Omega_t^{-1} \vec \alpha~,
\ee
and with (\ref{mgele}), we eventually get
\be
\label{eq:panzgt}
\vec w_t = \frac{1}{\vec 1' \Omega_t^{-1} \vec \alpha}~ \Omega_t^{-1}
\vec \alpha.
\ee

By substitution in (\ref{eq:market}) and (\ref{eq:return}), we obtain
\be
\label{eq:uyezt}
\E \[r_m(t)\]=\frac{\vec \alpha' \Omega_t^{-1} \vec \alpha}{\(\vec 1 -
\vec \beta^0\)' \Omega_t^{-1} \vec \alpha} \quad {\rm and} \quad
\E \[\vec r_t\] = \vec \alpha + \frac{\vec {\alpha'} \Omega_t^{-1} \vec \alpha}
{\( \vec 1 - \vec \beta^0\)' \Omega_t^{-1} \vec \alpha}~ \vec \beta^0.
\ee

Remark that $\E\[\vec r_t\]$ is a deterministic function of time since
$\Omega_t$ is deterministic. In addition, substituting $\vec w_t$ in
(\ref{eq:fsjkg}) by its expression (\ref{eq:panzgt}), we show that
$\Sigma_t$ depends on $t$ through $\Omega_t$ only, and is therefore a
deterministic function of $t$. Therefore, both $\Sigma_t$ and $\E\[\vec r_t\]$
are deterministic, which justifies the use of the
results of Li and Ng (2000) concerning the optimal allocation strategy
in a dynamic mean-variance formulation.

Let us now define the vector of beta coefficients
\be
\label{eq:threj}
\vec \beta  = \frac{\Cov \(\vec r_t, r_m(t) \)}{\Var~ r_m(t)} =
\frac{\(\vec 1 - \vec \beta^0\)' \Omega_t^{-1} \vec \alpha}{\vec
\alpha' \Omega_t^{-1} \vec \alpha}~ \vec \alpha + \vec \beta^0~,
\ee
where the last equality results from the equations
(\ref{eq:market}-\ref{eq:return}). Then, using (\ref{eq:threj}),
the expressions (\ref{eq:uyezt}) yield
\be
\label{eq:pjzer}
\E \[ \vec r_t\] = \vec \beta \E\[r_m(t)\]~,
\ee
which is nothing but the fundamental CAPM prediction that
the excess return of each individual stock is
proportional to the excess return on the market portfolio. This shows that
the relation of the CAPM can be derived from the regression model (\ref{eq:capm})
together with the self-consistency condition (\ref{mvgmlsl}) under 
the assumption of the
existence of an equilibrium.


\pagebreak

\section*{Appendix B:  observable beta's and ortho-normality
conditions in one-factor models}

We start with the decomposition (\ref{eq:rtyho}) of the vector of
disturbances $\vec \epsilon_t$ as the sum of a term proportional to $r_m(t)$ 
plus a contribution uncorrelated with $r_m(t)$.
We thus introduce two non-random vectors $\vec \delta $, $\vec \gamma$ and
the random vector $\vec u_t$, {\em uncorrelated} with $r_m(t)$ and
with zero mean
defined in (\ref{eq:rtyho}). The covariance matrix $\tilde \Omega_t$ of
$\vec u_t$ will be shown to be not full-rank in the following.

In order to express $\vec \delta$, $\vec \gamma$ and $\tilde \Omega_t$,
let us remark that
\be
\label{eq:iuetngf}
0 = \vec \delta + \vec \gamma \cdot \E \[r_m(t)\],
\ee
\be
\Omega_t = \Var~ r_m(t) \cdot \vec \gamma \vec \gamma' + \tilde \Omega_t,
\ee
and
\be
\vec \gamma = \frac{1}{\Var~ r_m(t)}~ \Cov \(\vec \epsilon_t, r_m(t) \).
\ee
Since
\be
\label{eq:ghdpo}
\Var~ r_m(t) = \frac{\vec \alpha' \Omega^{-1} \vec
\alpha}{\[\(\vec 1 - \vec \beta^0 \)'\Omega^{-1} \vec \alpha\]^2},
\quad {\rm and} \quad
\Cov \(\vec \epsilon_t, r_m(t) \) = \frac{\vec \alpha}{\(\vec 1 -
\vec \beta^0 \)'\Omega^{-1}
   \vec \alpha},
\ee
we obtain
\be
\label{eq:tyklh}
\vec \gamma = \frac{\(\vec 1 - \vec \beta^0\)' \Omega_t^{-1} \vec \alpha}{\vec
\alpha' \Omega_t^{-1} \vec \alpha}~ \vec \alpha
\ee
and
\be
\tilde \Omega_t = \Omega_t - \frac{1}{\vec \alpha' \Omega^{-1} \vec \alpha}
\vec \alpha \vec \alpha'.
\ee
It is straightforward to check that
\be
\label{eq:hdsqiu}
\tilde \Omega_t \( \Omega_t^{-1} \vec \alpha \) = 0
\ee
so that as asserted above, $\tilde \Omega_t$ is not full rank.

Beside, by (\ref{eq:uyezt}), (\ref{eq:iuetngf}) and (\ref{eq:tyklh}), we get
\be
\vec \delta = - \vec \alpha.
\ee

Now, substituting these relations into (\ref{eq:rtyho}) yields
\be
\vec \epsilon_t = -\vec \alpha + \frac{\(\vec 1 - \vec \beta^0\)'
\Omega_t^{-1} \vec \alpha}{\vec
\alpha' \Omega_t^{-1} \vec \alpha}~ \vec \alpha \cdot r_m(t)  + \vec u_t,
\ee
and replacing $\vec \epsilon_t$ into (\ref{eq:capm}), we get
\bea
\vec r_t &=& \[\frac{\(\vec 1 - \vec \beta^0\)' \Omega_t^{-1} \vec \alpha}{\vec
\alpha' \Omega_t^{-1} \vec \alpha}~ \vec \alpha + \vec \beta^0\]
\cdot r_m(t) + \vec u_t,\\
&=& \vec \beta \cdot r_m(t) + \vec u_t.
\label{eq:eriugh}
\eea
The $\alpha$ terms have disappeared as a direct consequence of endogeneity,
that is, the market portfolio is expressed through (\ref{mvgmlsl})
in terms of the basket of assets it is supposed to explain.

In the present form, the model is self-consistent. Indeed, left
multiplying the last equation of (\ref{eq:eriugh})
by $\vec w_t'$ yields
\be
r_m(t) = \underbrace{\vec w_t' \vec \beta}_{=1} \cdot r_m(t) + \vec
w_t' \vec u_t,
\ee
so that
\be
\label{eq:vsdhg}
\vec w_t' \vec u_t = 0,
\ee
which is consistent with the fact that $\tilde \Omega_t$ is not full
rank. Indeed, taking the variance of both side of the equation (\ref{eq:vsdhg})
leads to
\be
\vec w_t' \underbrace{\tilde \Omega_t \vec w_t}_{=0~
{\rm by~(\ref{eq:panzgt})~and~(\ref{eq:hdsqiu})}} = 0
\ee

These calculations show that, thanks to
(\ref{eq:eriugh}) and {\em under the assumption that $r_m(t)$ is
observable}, the OLS estimator of (\ref{eq:capm}) provides an estimate
of $\vec \beta$ (and not $\vec \beta^0$ and $\vec \alpha$ which
are unobservable). This comes with two conditions of consistency, $\vec w_t'
\vec \beta =1$ and $w_t' \vec u_t = 0$.


\pagebreak

\section*{Appendix C:  Breakdown of the main CAPM prediction when the proxy
is not the true market portfolio in the one-factor model}

Let us investigate the impact of replacing the market factor by a
proxy and derive expression (\ref{mgmlss}).
Let us denote by $\tilde w$ the portfolio of the market proxy.
Left multiplying (\ref{eq:eriugh2}) by $\tilde w'$ we get
\be
\tilde r_t \stackrel{def}{=} \tilde w' \vec r_t = \tilde w' \vec
\beta \cdot r_m(t) +
   \tilde w' \vec u_t,
\ee
which allows us to express $r_m$ as a function of $\tilde r$ (provided
that $\tilde w' \vec \beta \neq 0$) and, by (\ref{eq:eriugh}), we have
\be
\label{eq:usdqd}
\vec r_t = \frac{1}{\tilde w' \vec \beta} \vec \beta \cdot
\tilde r_t + \underbrace{\( {\rm Id} -\frac{\vec \beta \tilde
w'}{\tilde w' \vec \beta}\)
\vec u_t}_{\vec v_t}~.
\ee
Again, the residual vector $\vec v_t$ is correlated with $\tilde
r_t$, which implies
that such a model cannot be directly estimated by the OLS estimator.

Performing a decomposition similar to (\ref{eq:rtyho}),
we define another residual vector $\vec \eta_t$ such that
\be
\label{eq:zrce}
\vec v_t = \vec \gamma \(\tilde r_t - \E\[ \tilde r_t \] \) + \vec \eta,
\ee
with $\Cov\(\vec \eta ,\tilde r_t\) = 0$ and $\E\[\vec \eta \]=0$, by
construction.

Following the same lines of reasoning as in Appendix B, we can
express $\vec \gamma$ and the
covariance matrix of $\vec \eta$. We find
\be
\vec \gamma = \frac{\tilde \Omega \tilde w - \(\tilde w' \tilde \Omega
\tilde w \) \frac{\vec \beta}{\tilde w'\vec \beta}}{\(\tilde w' \tilde
\Omega \tilde w \) + \(\tilde w' \vec \beta\)^2 \Var~r_m},
\ee
where $\tilde \Omega$ is the covariance matrix of $\vec u_t$.
Thus, by (\ref{eq:usdqd}) and (\ref{eq:zrce}), we obtain
\be
\label{eq:dhml}
\vec r_t = \frac{1}{\tilde w' \vec \beta} \vec \beta \cdot \E[\tilde
r_t] + \underbrace{\frac{\tilde \Omega \tilde w + \(\tilde w' \vec
\beta\) \Var~r_m ~ \vec \beta}{\(\tilde w' \tilde \Omega \tilde w \) +
\(\tilde w' \vec \beta\)^2 \Var~r_m}}_{\vec{\tilde \beta}}~ \(\tilde r_t
- \E\[ \tilde r_t \] \) + \vec \eta~,
\ee
or, equivalently
\bea
\label{eq:qfdht}
\vec r_t &=& \underbrace{\( \frac{1}{\tilde w' \vec \beta} \vec \beta
- \vec {\tilde \beta}\)}_{-\vec \gamma} \cdot \E[\tilde r_t] + \vec
{\tilde \beta}
   \cdot \tilde r_t + \vec \eta,\\
&=& \( \E\[r_m\] \vec \beta -  \E\[\tilde r_t\] \vec {\tilde \beta}\)
+ \vec {\tilde \beta} \cdot \tilde r_t + \vec \eta~,
\eea
which is the announced result (\ref{mgmlss}). Thus,
using a proxy different from the true market portfolio
yields a non-vanishing intercept in the regression
of asset returns as a function of the proxy returns.

Left multiplying $\vec{\tilde \beta}$ in (\ref{eq:dhml}) by $\tilde w'$
yields $\tilde w' \vec{\tilde \beta} = 1$, which is the usual
normalization condition. Then, left multiplying the intercept in
(\ref{eq:qfdht}) by $\tilde w'$, we obtain
\be
\tilde w ' \cdot \( \frac{1}{\tilde w' \vec \beta} \vec \beta
- \vec {\tilde \beta}\) =0,
\ee
which provides a new orthogonality condition. Obviously, left
multiplying (\ref{eq:qfdht}) by $\tilde w'$ and accounting for the two
previous constraints leads to $\tilde w' \eta_t =0$.

To sum up, when dealing with a proxy of the market portfolio, the
self-consistency conditions lead us to cast the CAPM into a statistical
regression model with a non-vanishing intercept and the regression
has to obey three constraints on
the parameter and the residuals of the regression, namely a
normalization condition
\be
\tilde w' \vec{\tilde \beta} = 1~,
\ee
and two orthogonality conditions
\be
\tilde w' \eta_t =0 \quad {\rm and} \quad \tilde w ' \cdot \(
\frac{1}{\tilde w' \vec \beta} \vec \beta
- \vec {\tilde \beta}\) =0~.
\ee


\pagebreak

\section*{Appendix D: Breakdown of the main CAPM prediction and normality
condition when the proxy
is not the true market portfolio in the multi-factor model}

We derive the result announced in section \ref{mnvgkls}, in the case
where the individual asset
returns can be explained by {\em exactly} $q$ factors.
Then, $q$ factor proxies
can be built by defining $q$ non-degenerate portfolios of the traded
assets. Let us
denote by $\tilde W$ the matrix whose columns represent the $q$
portfolios and by $\vec v_t$ the vector of the $q$ proxies. By equation
(\ref{eq:kukggf}), we have
\be
\label{eq:hfdi}
\tilde W' \vec r_t = \vec v_t = \tilde W' B \vec u_t + \tilde W' \vec
\epsilon_t,
\ee
and, assuming that the matrix $\tilde W' B$ is full rank, we obtain
\be
\vec u_t = \( \tilde W' B\)^{-1} \vec v_t - \( \tilde W' B\)^{-1}
\tilde W' \vec \epsilon_t,
\ee
so that (\ref{eq:kukggf}) can be rewritten as
\be
\vec r_t = B \(\tilde W' B\)^{-1} \vec v_t + \underbrace{\[ {\rm Id} - B
\( \tilde W' B\)^{-1} \tilde W' \] \vec \epsilon_t}_{\vec \eta_t}~,
\ee
where the disturbance $\vec \eta_t$ is correlated with $\vec v_t$, since
both $\vec v_t$ and $\vec \eta_t$ depend on $\vec \epsilon_t$.

As in Appendix B, we can define a new residual vector $\vec \nu_t$ such that
\be
\vec \eta= \Gamma \cdot \(\vec v_t  - \E \[\vec v_t \]\) + \vec \nu_t
\ee
with $\E\[\vec \nu_t\]=0$ and $\Cov\(\vec v_t, \vec \nu_t\)=0$. As
usual, we obtain
\be
\label{eq:cxku}
\Gamma = \Cov\(\vec \eta_t, \vec v_t \) \cdot \Cov\(\vec v_t, \vec v_t \)^{-1},
\ee
with
\be
\label{eq:hpjo}
\Cov\(\vec \eta_t, \vec v_t \) = \[ {\rm Id} - B \( \tilde W' B\)^{-1}
\tilde W' \] \Omega_t \tilde W,
\ee
and
\be
\Cov\(\vec v_t, \vec v_t \) = \tilde W' \Omega_t \tilde W + \( \tilde W'
B \) \Cov \(\vec u_t, \vec u_t \) \( B' \tilde W\)~,
\ee
where $\Omega_t = \Cov\(\vec \epsilon_t, \vec \epsilon_t \)$. Finally, one gets
\bea
\label{eq:pdfig}
\vec r_t &=& \underbrace{B \(\tilde W' B \)^{-1} \cdot \E \[ \vec v_t \]}_{=B
\E \[\vec u_t\]=\E\[\vec r_t\],~~ {\rm by~ (\ref{eq:hfdi})}} + \[ \Gamma + B
   \(\tilde W' B \)^{-1} \] \cdot \( \vec v_t - \E \[ \vec v_t \] \) +
\vec \nu_t,\\
&=& \underbrace{- \Gamma \cdot \E \[\vec v_t \]}_{\vec \alpha} + \underbrace{\[
\Gamma + B \(\tilde W' B \)^{-1} \]}_{\tilde B} \vec v_t + \vec
\nu_t. \label{eq:posoir}
\eea
When using a set of $q$ factor proxies instead of the true set of risk factors,
we find that a non-zero intercept $\vec \alpha$ appears as for the
one-factor case, and the normalization and orthogonality relations
\be
\label{eq:pdgho}
\tilde W' \tilde B = {\rm Id}, \quad \tilde W' \alpha = \vec 0, \quad
{\rm and}, \quad \tilde W' \vec \nu_t = \vec 0,
\ee
still hold since equation (\ref{eq:hpjo}) implies $\tilde W' \Cov\(\vec \eta_t,
\vec v_t \) = 0$, which yields $\tilde W' \Gamma =0$, by (\ref{eq:cxku}).


\pagebreak

\section*{Appendix E: Analysis of the case where the number $r$ of
factor proxies is larger than the number $q$ of true factors}

In this case, equation (\ref{eq:hfdi}) still holds, but the matrix
$\tilde W' B$ is
not a $q\times q$ matrix anymore, it is a $r \times q$ matrix, where
$r>q$ is the number of chosen factors. As a consequence,
$\(\tilde W' B\)^{-1}$ does not exist and has to be replaced by its
(left) pseudo-inverse
\be
\(\tilde W' B\)^{-1} \longrightarrow \(\tilde W' B\)^\dagger
\stackrel{def}{=} \(B' \tilde W \tilde W' B\)^{-1} B' \tilde W~,
\ee
which is such that $\(\tilde W' B\)^\dagger \cdot \(\tilde W' B\) = {\rm
Id}_q$. Since the calculation performed in Appendix D up to equation
(\ref{eq:posoir}) involves only the (left) inversion of the matrix $\tilde
W' B$, it remains valid if we
replace $\(\tilde W' B\)^{-1}$ by $\(\tilde W'
B\)^\dagger$, so that $\tilde B$ becomes
\bea
\tilde B &=&  \Gamma + B \(\tilde W' B \)^\dagger,  \label{mgbmlsls}\\
&=& \Gamma + B \(B' \tilde W \tilde W' B\)^{-1} B' \tilde W,
\eea
while the new expression of the intercept is
\be
\vec \alpha = \[ {\rm Id} - \Gamma \tilde W' - B \( \tilde W'
B\)^\dagger \tilde W' \] \cdot \E \[\vec r_t \].
\ee
Note that the existence of the inverse of $\Cov\( \vec v_t, \vec v_t \)$ is
ensured as it does not require the invertibility of $\tilde W'B$, and
thus $\Gamma$ is well defined.

The orthogonality and normalization conditions 
(\ref{eq:pdgho2},\ref{eq:pdgho}) still
hold, but are slightly more difficult to derive. Indeed, from 
(\ref{mgbmlsls}), we have
\be
\label{eq:dfjkhf}
\tilde W' \tilde B = \tilde W'\Gamma + \tilde W' B \(\tilde W' B\)^\dagger,
\ee
with $\tilde W' B \(\tilde W' B\)^\dagger \neq {\rm Id}$ and $\tilde
W'\Gamma \neq 0$. Remarking that
\be
\tilde W' \cdot \Cov \(\vec \eta_t, \vec v_t \) = \[ {\rm Id} -
\underbrace{\tilde W' B \( \tilde W' B\)^\dagger}_{\neq {\rm Id}} \]
\tilde W' \Omega_t \tilde W,
\ee
and applying the matrix inversion formula (Golub and Van Loan, 1996), we get
\bea
\Cov\(\vec v_t, \vec v_t\)^{-1} &=& \(\tilde W' \Omega_t \tilde W\)^{-1}
\left\{ {\rm Id} - \(\tilde W' B\) \[ \Cov(\vec u_t,\vec
u_t)^{-1}\right. \right. \nonumber\\
&+& \left. \left. \(\tilde W' B\)'  \(\tilde W' \Omega_t \tilde W\)^{-1}
\(\tilde W' B\) \]\(\tilde W' B\)'  \(\tilde W' \Omega_t \tilde W\)^{-1}
\right\},
\eea
so that
\bea
\tilde W' \Gamma &=& \[ {\rm Id} - \tilde W' B \( \tilde W' B\)^\dagger
\] \cdot \left\{ {\rm Id} - \(\tilde W' B\) \[ \Cov(\vec u_t,\vec
u_t)^{-1}\right. \right.\nonumber\\
   &+& \left. \left.  \(\tilde W' B\)'  \(\tilde W' \Omega_t \tilde
   W\)^{-1} \(\tilde W' B\) \]\(\tilde W' B\)'  \(\tilde W' \Omega_t
   \tilde W\)^{-1} \right\},\\ &=& {\rm Id} - \tilde W' B \(\tilde W'
   B\)^\dagger~,
\eea
and thus, by (\ref{eq:dfjkhf}),  we recover (\ref{eq:pdgho}).


\pagebreak

\section*{Appendix F: Derivation of the procedure to calibrate
the self-consistent multi-factor model}

We start with the multi-factor model (\ref{eq:kukggf})
\bea
\vec r_t &=& B \vec u_t + \vec \epsilon_t\\
&=& B W' \vec r_t + \epsilon_t~,
\eea
and the properties (\ref{mvmb},\ref{nbjklls;},\ref{mgmlsww}).
Our goal is to obtain the procedure summarized in
section \ref{nvbkdd} to estimate the beta's $B$ and the $Q$ replicating
portfolio weights $W=\(W_1, W_2, \ldots, W_Q\)$ by a suitable
calibration of the model to the data consisting of the returns
of the $N$ assets of the market over a given period of time of length $T$.

Given the above properties (\ref{mvmb},\ref{nbjklls;},\ref{mgmlsww}), the
estimation of the multi-factor model
amounts to finding the $N \times Q$ matrices $B$ and $W$ which minimize
\be
\label{eq:dfkh}
\sum_{t=1}^T \vec r_t' \( {\rm Id} - BW'\)' \( {\rm Id} - BW'\) \vec r_t
\ee
under the constraints (\ref{mvmb},\ref{nbjklls;}).
The last condition (\ref{mgmlsww}) $(W' \vec \epsilon_t = 0)$ is
automatically fulfilled
by the least square regression.

Introducing the $Q \times 1$ vector $2\vec \mu$ and the $Q \times
Q$ matrix $2\Lambda$ of Lagrange multipliers, the Lagrangian of the
system reads
\be
L(W,B)= r_{it}r_{it} - 2 \cdot r_{it}B_{ik}W_{lk}r_{lt} +
r_{it}W_{ik}B_{jk}B_{jm}W_{lm}r_{lt} - 2\mu_j \( \sum_i W_{ij}-1\) -
2\lambda_{ij}\( B_{ki} W_{kj} - \delta_{ij}\)~,   \label{nbmls;}
\ee
where we sum over repeated subscripts.

Differentiating $L(W,B)$ with respect to $B_{rs}$ and $W_{rs}$
respectively yields
\bea
\partial_{B_{rs}} L &=& r_{it} W_{is} B_{rm} W_{lm} r_{lt} - r_{rt}
W_{ls} r_{lt} - \lambda_{sj} W_{rj},\\
\partial_{W_{rs}} L &=& r_{rt} B_{js} B_{jm} W_{lm} r_{lt} - r_{it}
B_{is} r_{rt} - \mu_s -\lambda_{is} B_{ri}~.
\eea
The minimization of (\ref{eq:dfkh}) thus leads to the first order condition
\bea
BW'\vec r_t \vec r_t' W -  \vec r_t \vec r_t' W - W \Lambda' &=& 0,\\
\vec r_t \vec r_t' W B' B - \vec r_t \vec r_t' B - \vec 1_N \vec \mu'
- B \Lambda &=& 0.
\eea

Summing up, we have to find two $N \times Q$ matrices $B$ and $W$, a $Q
\times Q$ matrix $\Lambda$ and a $Q \times 1$ vector $\vec \mu$ solution of
\bea
\label{eq1}&&BW'\vec r_t \vec r_t' W -  \vec r_t \vec r_t' W - W
\Lambda' = 0,\\
\label{eq2}&&\vec r_t \vec r_t' W B' B - \vec r_t \vec r_t' B - \vec 1_N
\vec \mu' - B \Lambda = 0,\\
\label{eq3}&&W' \vec 1_N = \vec 1_Q,\\
\label{eq4}&&W' B = {\rm Id}_Q.
\eea

We can simplify this program using the following manipulations.
\bea
\label{eq:jf}
W' \cdot (\ref{eq2}) \quad &\stackrel{(\ref{eq4})}{\Longrightarrow}& \quad
\Lambda = W' \vec r_t \vec r_t' W B'B - W' \vec r_t \vec r_t' B' -
\underbrace{W' \vec 1_N}_{=\vec 1_Q~ (\ref{eq3})} \vec \mu'\\
B' \cdot (\ref{eq1}) \quad &\stackrel{(\ref{eq4})}{\Longrightarrow}& \quad
\Lambda' = B' B W' \vec r_t' \vec r_t W - B' \vec r_t \vec r_t' W\\
&\Longrightarrow& \quad
\Lambda = W' \vec r_t \vec r_t' W B'B - W' \vec r_t \vec r_t' B',\label{eq:yju}
\eea
Therefore, by (\ref{eq:jf}) and (\ref{eq:yju}), we must have
\be
\vec 1_Q \cdot \vec \mu' = 0
\ee
which leads to
\be
\vec \mu = 0~.    \label{mhmltl}
\ee
and equation (\ref{eq2}) simplifies into
\be
\vec r_t \vec r_t' W B' B - \vec r_t \vec r_t' B - B \Lambda = 0. \label{eq2'}
\ee

Then,
\bea
W' \cdot (\ref{eq1}) \quad &{\Longrightarrow}& \quad
\underbrace{W' B}_{{\rm Id}~(\ref{eq4})} W' \vec r_t' \vec r_t W - W'
\vec r_t \vec r_t' W = W'W \Lambda'\\
&\Longrightarrow& \quad
W' W \Lambda' =0,
\eea
so that
\be
\Lambda=0 ~,    \label{mhmmj;mn;t}
\ee
since the matrix $\[W'W \]$ is full rank (equal to $Q$) and therefore
invertible.

The equations (\ref{eq1},\ref{eq2}) reduce to
\bea
&&BW'\vec r_t \vec r_t' W -  \vec r_t \vec r_t' W  = 0,\\
\label{eq2''}&&\vec r_t \vec r_t' W B' B - \vec r_t \vec r_t' B  = 0,
\eea

Since $[\vec r_t \vec r_t']$ is invertible, provided that $T>N$,
equation (\ref{eq2''}) leads to $W B' B - B  = 0$, and finally
\be
W= B \[B'B\]^{-1}~,   \label{wpoqpuvqa;nqf}
\ee
since the matrix $\[B'B \]$ is full rank (equal to $Q$) and therefore
invertible.

Note that (\ref{eq4}) is automatically satisfied, so that the search
for $B$ and $W$ in the system
(\ref{eq1}-\ref{eq4}) reduces to finding a matrix $B$ such that
\bea
&&BW'\vec r_t \vec r_t' W -  \vec r_t \vec r_t' W  = 0,\\
&& W= B \[B'B\]^{-1}\\
&&W' \vec 1_N = \vec 1_Q,
\eea
or equivalently by using (\ref{wpoqpuvqa;nqf})
\bea
&&B[B'B]^{-1}B'\vec r_t \vec r_t' B =  \vec r_t \vec r_t' B,   \label{eqA}\\
&& W= B \[B'B\]^{-1}~,   \label{eqB} \\
&&B'\[ \vec 1_N - B \vec 1_Q\]=0~.   \label{eqC}
\eea

Finding the solution of this system above is not straightforward.
An alternative approach is to go back to the quadratic form (\ref{eq:dfkh})
to minimize, and use (\ref{wpoqpuvqa;nqf}) to replace
$W$ by $B\[B'B\]^{-1}$. In addition, from (\ref{mhmltl}) and
(\ref{mhmmj;mn;t}),
we know now that both $\vec \mu$ and $\Lambda$ are zero. Then, the
optimal matrix $B$ we are looking for is solution of
\be
\min_B \sum_t \vec r_t' \( {\rm Id} - B \[B' B\]^{-1}B'\) \vec r_t
\ee
under the constraint
\be
B'\[ \vec 1_N - B \vec 1_Q\]=0~.   \label{mbhmlr}
\ee
This minimization has the same solution as
\bea
\max_B \sum_t \vec r_t'  B \[B' B\]^{-1}B' \vec r_t &=& \max_B {\rm Tr}
\( R'B \[B' B\]^{-1}B' R\),\\
&=&\max_B {\rm Tr} \( B' R R'B \[B' B\]^{-1}\)\\
\eea
under the same constraint (\ref{mbhmlr}),
where $R$ the $N \times T$ matrix $R= \(\vec r_1, \vec r_2, \ldots,
\vec r_T \)$.
$T$ is the duration of the time interval over which the data is available.

Using the transformation
\be
RR'=P'DP~,    \label{mbm,;w;w}
\ee
where $D$ is a diagonal matrix and $P$ is the matrix of the (orthogonal)
eigenvectors of $RR'$, we thus have to solve
\be
\max_{\tilde B} {\rm Tr} \( \tilde B' D \tilde B \[\tilde B' \tilde B\]^{-1}\)
\ee
under the constraint
\be
\label{eq:QSG}
\tilde B'P \vec 1_N =  \tilde B' \tilde B \vec 1_Q,
\ee
where $\tilde B = P B$ is a full rank $N \times Q$ matrix.

Any $N \times Q$ ($N \ge Q)$ matrix $\tilde B$ admits a
singular value decomposition
\be
\tilde B = U \nabla V',
\label{mgvjslw}
\ee
where $U$ is an $N \times Q$ matrix, $\nabla$ and $V$ are $Q \times Q$
matrices with $\nabla$ diagonal and
\bea
U' U &=& {\rm Id}_Q,\\
V' V &=& V V' = {\rm Id}_N~.
\eea
With the singular value decomposition (\ref{mgvjslw}), the constraint
(\ref{eq:QSG}) becomes
\be
U' P \vec 1_N = \nabla V' \vec 1_Q.
\ee
Thus, defining $\nabla$ as the diagonal matrix whose i$^{th}$ diagonal
element $\nabla_i$ is given by the ratio of the i$^{th}$ component of the
vector $U' P \vec 1_N$ over the i$^{th}$ component of the vector $V' \vec 1_Q$
\be
\label{eq:dksfuh}
\nabla_i = \frac{\[U' P \vec 1_N\]_i}{\[V' \vec 1_Q\]_i}~,
\ee
any matrix $\tilde B$ solution of the constraint (\ref{eq:QSG}) can
be written as
\be
\tilde B = U \nabla V'~,
\ee
where
\begin{itemize}
\item $V$ is any $Q \times Q$ orthogonal matrix whose lines add up to a
non vanishing constant (to ensure the existence of $\nabla_i$ in
(\ref{eq:dksfuh}), for all $i=1, \ldots,Q$),

\item $\nabla$ is the $Q \times Q$ diagonal matrix defined by
(\ref{eq:dksfuh}),

\item and $U$ is any $N \times Q$ matrix such that $U'U = {\rm Id}_Q$.
\end{itemize}

By circular permutation
\be
{\rm Tr} \( \tilde B' D \tilde B \[\tilde B' \tilde B\]^{-1}\) = {\rm
Tr} \( D \tilde B \[\tilde B' \tilde B\]^{-1} \tilde B'\)~,
\ee
and since a straightforward  calculation shows that
\be
\tilde B \[\tilde B' \tilde B \]^{-1} \tilde B' = U U'~,
\ee
our maximization program becomes
\be
\max_U {\rm Tr} \( D U U' \) = \max_U {\rm Tr} \( U' D U \),
\ee
under the constraint
\be
U'U = {\rm Id}_Q.
\ee

Recalling that $D$ is the diagonal matrix which elements equal to the
eigenvalues of the matrix $RR'$ and assuming that these eigenvalues are
sorted in decreasing order $(D_{11} \ge D_{22} \ge \cdots \ge D_{NN})$,
the simplest solution of the maximization program is
\be
U= \(
\begin{array}{c}
{\rm Id}_Q\\
0
\end{array}\)~.   \label{m,h,a;,s}
\ee

To sum up, the set of optimal solutions of the original problem
(\ref{eq:dfkh}) is
given by
\bea
B &=& P' U \nabla V',   \label{mnbkl}\\
W &=& P' U \nabla^{-1} V'~.   \label{bnjewla}
\eea
While $P$ is unique by the decomposition (\ref{mbm,;w;w})
and $U$ is also fixed to (\ref{m,h,a;,s}), the matrix $V$ can be
any $Q \times Q$ orthogonal matrix whose lines add up to a
non vanishing constant. This expresses simply the rotational degeneracy
of the $Q$ factors.

\pagebreak


{\bf References}

Alexander, G.J. and A.M. Baptista (2002)  Economic
implications of using a mean-VaR model for portfolio selection: A
comparison with mean-variance analysis, Journal of Economic
Dynamics \& Control 26, 1159--1193.

Axtell, R.L. (2001) Zipf distribution of U.S. firm sizes, Science 293, 1818-1820.

Bai, J. and S. Ng, S. (2002)
Determining the Number of Factors in Approximate Factor Models,
Econometrica 70 (1), 191-221.

Black, F. (1972) Capital market equilibrium with restricted borrowing, 
Journal of Business 45 (July), 444-455.

Blume, M. E. (1971) On the Assessment of Risk, Journal of Finance, 26, 1-10.

Blume, M. E. (1975) Betas and their Regression Tendencies, Journal of Finance, 30, 785-795.

Bodie, Z., A. Kane, A.J. Marcus (2004) Investments, 6th ed., McGraw-Hill/Irwin.

Bos, T. and Newbold, P. (1984) An Empirical Investigation of the Possibility of Stochastic
Systematic Risk in the Market Model, Journal of Business, 57, 35-41.

Broomhead, D.S., King G. (1986) Extracting qualitative dynamics from experimental
data, Physica D 20, 217-236.

Chan, K.S., Tong H. (2001) Chaos: A statistic perspective, Spring-Verlag, New York.

Collins, D. W. Ledolter, J. and Rayburn, J. (1987) Some Further Evidence on the Stochastic
Properties of Systematic Risk, Journal of Business, 60, 425-448.

Connor, G. and R. Korajzcyk (1993) A test for the number of factors in an approximate
factor model, Journal of Finance 48 (4), 1263-1291. 

Dunteman, G.H. (1989) Principal components analysis,
Sage Publication, California.

Fama, E.F. and K.R. French (1992) The Cross-Section of Expected
Stock Returns, Journal of Finance 47,~427--465.

Fama, E.F. and K.R. French (1993) Common Risk Factors in
the Returns on Stocks and Bonds, Journal of Financial Economics 33, 3--56.

Fama, E.F. and K.R. French (1995) Size and Book-to-Market
Factors in Earnings and Returns, Journal of Finance 50, 131--155.

Fama, E.F. and K.R. French (2004)
The CAPM: Theory and Evidence,  Journal of Economic Perspectives 18
(August), 25-46.

Fang, H.B. and T. Lai (1997)
Co-kurtosis and capital asset pricing, Financial Review 32,~293--307.

Figlewski, S. (2004) Forecasting volatility, Blackwell.

Golub, G.H. and C.F. Van Loan (1996) {\it Matrix Computations},
3rd edition (John Hopkins: Baltimore, MD), p51.

Greene, W. (2003) Econometric Analysis, 5th Edition, Prentice Hall.

Hakansson, N.H. (1971) On optimal myopic portfolio policies, with and
without serial correlation of yields, Journal of Business 44, 324--334.

Harvey, C.R. and A. Siddique (2000)
Conditional skewness in asset pricing tests,
Journal of Finance 55, 1263--1295.

Hwang, S. and S. Satchell (1999)
Modelling emerging market risk premia using higher moments,
International Journal of Finance \& Economics 4,~271--296.

IMF, International Monetary Fund (2004) Compilation Guide
on Financial Soundness Indicators, IMF, Washington DC, Appendix VII, Glossary.

Jolliffe, I.T. (2002) Principal Component Analysis, 2nd ed.,
Springer Series in Statistics, New York.

Krauss, A. and R. Litzenberger (1976)
Skewness preference and the valuation of risk assets
Journal of Finance 31, 1085--1099.

Lee, C.F. and Chen, CR. (1982) Beta Stability and Tendency: An Application of the Variable
Mean Response Regression Model, Journal of Economics and Business, 34, 201-206.

Li, D. and W.L. Ng (2000) Optimal dynamic portfolio selection:
multiperiod mean-variance formulation, Mathematical Finance 10, 387--406.

Lifshitz, E.M., L.P. Pitaevskii and V.B. Berestetskii (1982)
Quantum Electrodynamics, 2nd ed., Vol. 4,
Butterworth-Heinemann.

Lim, K.G. (1989) A new test for the
three-moment capital asset pricing model, Journal of
Financial \& Quantitative Analysis 24, 205--216.

Loretan, M. (1997) Generating market risk scenarios using principal components analysis: 
Methodological and practical considerations, In The Measurement of Aggregate Market Risk, CGFS
Publications No. 7, pages 23-60. Bank for International Settlements, November 1997
(Available at http://www.bis.org/publ/ecsc07.htm).

Malevergne, Y. and D. Sornette (2006a)
Multi-Moments Method for Portfolio Management:
Generalized Capital Asset Pricing Model in
Homogeneous and Heterogeneous markets, in B. Maillet and E. Jurczenko (eds.): Multi-moment Asset Allocation and Pricing Models (Wiley \& Sons), pp. 165-193.

Malevergne, Y. and D. Sornette (2006b)
New Irreducible Non-Diversifiable Risks in Asset Pricing Models,
working paper, ETH Zurich.

Michaud, R.A. (2003) A practical framework for portfolio choice,
Journal of Investment Management 1 (2), 1-16.

Ohlson, J. and Rosenberg, B. (1982) Systematic Risk of the CRSP Equal-weighted Common
Stock Index: A History Estimated by Stochastic Parameter Regression, Journal of Business, 55,
121-145.

Petkova, R. (2006) Do the Fama-French factors proxy
for innovations in predictive variables? Journal of Finance,
forthcoming (April).

Pliska, S.R. (1997) {\it Introduction to Mathematical Finance}. Blackwell: Malden, MA.

Polimenis, V. (2005) On the concavity of jump equity premia, Finance
Letters 3 (1), paper 18, February.

Roll, R. (1977) A critique of the asset pricing theory's tests, 
Part I: On past and potential testability of the theory,
Journal of Financial Economics 4, 129-176.

Roll, R. (1994)
What every CFO should know about scientific progress in financial
economics: What is known and what remains to be resolved.
 Financial Management 23(2),~69--75.

Roll, R. and S.A. Ross (1984) The arbitrage pricing theory approach to
strategic portfolio planning, Financial Analysts Journal (May/June), 14-26.

Ross, S.A. (1976) The Arbitrage Theory of Capital Asset Pricing, Journal of
Economic Theory (December), 341-60.

Rubinstein, M. (1973) The fundamental theorem
of parameter-preference security valuation, Journal of Financial \&
Quantitative Analysis 8,~61--69.

Samuelson, P.A. (1969) Lifetime portfolio selection by dynamic
stochastic programming, The Review of Economics and Statistics 50, 239--246.

Satchell, S. and J. Knight (2002) Forecasting Volatility in the Financial Markets, 
2nd ed. (Quantitative Finance), Butterworth-Heinemann.

Sharpe, W.F. (1990) Capital asset prices with and without negative
holdings, Nobel Lecture, December 7, 1990.

Simmonds, R., La Motte, L. and McWhorter, A. (1986) Testing for Nonstationarity of Market
Risk: An Exact Test and Power Considerations, Journal of Financial and Quantitative Analysis,
21, 209-220.

Vautard, R., Yiou P., Ghil M. (1992) Singular-spectrum analysis: A toolkit for short,
noisy chaotic signals, Physica D 58, 95-126.


\clearpage

\clearpage

\clearpage

\clearpage

\clearpage

\clearpage

\clearpage

\clearpage

\end{document}